\documentclass[12pt]{article}
\usepackage[dvips]{graphicx}
\usepackage{amsmath}
\usepackage{graphicx}
\usepackage{amsfonts}
\usepackage{amssymb}
\usepackage{epsfig}
\usepackage[usenames]{color}

\newcommand{\be}{\begin{equation}}
\newcommand{\ee}{\end{equation}}
\newcommand{\bea}{\begin{eqnarray}}
\newcommand{\eea}{\end{eqnarray}}
\newcommand{\ba}{\begin{array}}
\newcommand{\ea}{\end{array}}
\newcommand{\beas}{\begin{eqnarray*}}
\newcommand{\eeas}{\end{eqnarray*}}
\newcommand{\bes}{\begin{equation*}}
\newcommand{\ees}{\end{equation*}}

\def\i2           {\mbox{$\frac{i}{2}$}}

\def\del           {\delta}

\begin{document}
\title{\bf Thermodynamic Properties of Holographic Multiquark and the Multiquark Star }

\author{P. Burikham$^{1,}$$^{2}$\thanks{Email:piyabut@gmail.com, piyabut.b@chula.ac.th}, E. Hirunsirisawat$^{1}$\thanks{Email:ekapong.hirun@gmail.com,
ekapong.h@student.chula.ac.th},S.
Pinkanjanarod$^{1,}$$^{3}$\thanks{Email:quazact@gmail.com, Sitthichai.P@student.chula.ac.th}\\$^1$ {\small {\em  Theoretical High-Energy Physics and Cosmology Group, Department of Physics,}}\\
{\small {\em Faculty of Science, Chulalongkorn University, Bangkok
10330, Thailand}}\\$^{2}$ {\small {\em  Thailand Center of
Excellence in Physics, Ministry of Education, Bangkok,
Thailand}}\\$^{3}$ {\small {\em Department of Physics, Faculty of
Science, Kasetsart University, Bangkok 10900, Thailand}}}

\maketitle

\begin{abstract}
\noindent  \\

    We study thermodynamic properties of the multiquark nuclear
matter.  The dependence of the equation of state on the colour
charges is explored both analytically and numerically in the
limits where the baryon density is small and large at fixed
temperature between the gluon deconfinement and chiral symmetry
restoration.  The gravitational stability of the hypothetical
multiquark stars are discussed using the
Tolman-Oppenheimer-Volkoff equation.  Since the equations of state
of the multiquarks can be well approximated by different power
laws for small and large density, the content of the multiquark
stars has the core and crust structure.  We found that most of the
mass of the star comes from the crust region where the density is
relatively small.  The mass limit of the multiquark star is
determined as well as its relation to the star radius.  For
typical energy density scale of $10~ \text{GeV}/\text{fm}^{3}$,
the converging mass and radius of the hypothetical multiquark star
in the limit of large central density are approximately $2.6-3.9$
solar mass and 15-27 km. The adiabatic index and sound speed
distributions of the multiquark matter in the star are also
calculated and discussed. The sound speed never exceeds the speed
of light and the multiquark matters are thus compressible even at
high density and pressure.

\end{abstract}

\newpage
\section{Introduction}

All of the high energy experiments which fail to produce a free
quark are strong evidences that the coupling constant of the
strong interaction becomes nonperturbatively large at low energy
and large distance.  Quarks and gluons are said to be confined
within hadrons and the colourless condition becomes a requirement
of an assembly of quarks at low energy. However, when the energy
or temperature scale of a system of quarks and gluons increases,
the coupling of the strong interaction tends to be weaker and
finally we expect the deconfinement to occur.  In addition, if the
quarks and gluons are compressed extremely tightly together,
quarks could interact with neighbouring quarks and gluons equally
and become effectively deconfined from the mesonic or baryonic
bound state. In the latter case, the coupling could still be
strong despite of the deconfinement.  Nevertheless, we could also
have the situation where gluons are deconfined but the quarks are
not completely free due to the remaining Coulomb-type potential
from gluon exchanges between quarks.

Recently, the experimental results from collision of heavy ions
suggested that the nuclear deconfinement phase might have been
created in the laboratory and we might have produced the
quark-gluon plasma~(QGP).  The RHIC experiment revealed that the
produced QGP behaves like fluid with very small viscosity.
However, this property of small viscosity fluid is hard to be
understood in the picture of QGP as the gas of free quarks and
gluons. Additionally, lattice simulations show that QGP has
relatively high pressure right above the deconfinement temperature
$T_{c}$ which is again difficult to explain using the weakly
coupled quarks and gluons gas~\cite{kar,mp,shuryak_colored
states2}. It is possible that various coloured and colour-singlet
bound states of quarks and gluons could exist in the plasma at the
temperature $(1-3)T_c$~\cite{shuryak_colored
states1,shuryak_colored states2,shuryak_colored states3}.  The
existence of the coloured bound states could explain the problems
of high pressure, small viscosity, and the jet quenching of the
QGP at once.

Due to the large coupling of the strong interaction at low
energies, a perturbative method has limited applicability to the
high energy processes and phenomena.  The development of the
holographic principle and AdS/CFT correspondence~\cite{maldacena}
provides us with a new method to investigate the physics of
strongly coupled nuclear matter both in the low energy regime and
in the energy scale close to the deconfinement temperature.
Holographic models of meson were proposed by Juan Maldacena,
Soo-Jong Rey, Stefan Theisen, Jung-Tay
Yee~\cite{maldacena2,ry,rty}.  The Coulomb potential plus
screening effect of quark and antiquark are calculated from the
Nambu-Goto action of the string in the bulk spacetime at zero and
finite temperature.  For baryons, Witten, Gross and
Ooguri~\cite{witb,gross&ooguri} proposed a holographic baryon to
be a D-brane wrapping internal subspace of the background
spacetime with $N_{c}$ strings connected and stretching out to the
boundary.  For $AdS_{5}\times S^{5}$, the baryon vertex is a
D5-brane wrapping the $S^{5}$.  The basic requirement is that a
total of $N_{c}$ charges from the endpoint of the strings cancel
with the charge of the vertex itself.  A generalization of this
condition allows more strings to go in and come out of the vertex,
as long as the total charges from all of the string endpoints add
up to $N_{c}$~\cite{BISY,gho1,gho2,Car,Wen,bch}.  Baryon vertex
plus strings configuration in this case represent the holographic
multiquark states.  Generically they have colour charges but
because of the confinement, they can only exist in the deconfined
phase.

The coloured multiquark phase can be studied in the general
Sakai-Sugimoto model~(SS)\cite{ss lowE,ss more} in the
intermediate temperature above the gluon deconfinement but below
the chiral symmetry restoration temperature~\cite{Aharony}.  It
was found that the multiquark phase is thermodynamically stable
and preferred over the other phases in the gluon-deconfined plasma
provided that the density is sufficiently large~\cite{bch}.  The
situation of high density and moderate temperature could exist
inside certain classes of compact stars and it is thus interesting
to investigate the thermodynamical properties of the multiquark
nuclear matter as well as their contributions to the stability of
the dense stars. In this article, we will consider the
hypothetical multiquark star which obeys the equation of state
derived from the holographic multiquarks in the SS model.  With
the power-law approximation of the equations of state, we study
its gravitational stability using the Tolman-Oppenheimer-Volkoff
equation~(TOV)\cite{tov}.  The mass, density and pressure
distributions are obtained numerically.  The mass-radius relation
and the mass limit are also discussed.  Corresponding
hydrodynamical properties such as the sound speed of the
multiquark nuclear matter are explored within the star.  The
multiquark matters are found to be compressible throughout the
entire multiquark star.

This article is organized as the following.  Section 2 describes
the holographic setup for the multiquarks and the multiquark phase
in the gluon-deconfined SS model.  The thermodynamic relations and
the equations of state of the multiquark nuclear matter are
calculated and discussed in Section 3 and 4.  In Section 5, the
Einstein field equation for the spherically symmetric star is
solved to obtain the TOV equation.  Assuming the equations of
state derived in Section 3 and 4 for the multiquark nuclear
matter, we explore the gravitational physics of a hypothetical
multiquark star.  A mass-radius relation is derived and some
discussion on the more realistic situation is commented.  The
adiabatic index and the sound speed of the multiquark nuclear
matter within the star are studied.  Section 6 concludes the
article.

\section{Holographic multiquark configuration}

Since string theories in the bulk spacetime correspond to certain
gauge theories on the boundary of that space, it is natural to
find construction of the bound states of quarks in the form of
strings and branes.  While the meson is proposed to be the string
hanging in the bulk with both ends locating at the boundary of the
AdS space~\cite{maldacena2}, the baryon is proposed to be the
D$p$-brane wrapped on the $S^{p}$ with $N_c$ strings attached and
extending to the boundary of the bulk
space~\cite{witb,gross&ooguri}.

On the gauge theory side, hadrons exist in the confined phase as a
result of the linear part of the binding potential.  However, the
bound states of quarks can actually exist in the deconfined phase
at the intermediate temperatures above the deconfinement as well.
Even though gluons are free to propagate and the linear potential
is absent, the quarks can form bound state through the remaining
Coulomb-type potential due to the colour charges of the quarks.

 The holographic model of
non-singlet bound state was also proposed.  As is demonstrated in
Ref.~\cite{bch}, we can modify the Witten's baryon vertex by
attaching more strings to the vertex provided that the total
number of charges of all of the strings are preserved to $N_{c}$.
Some strings may extend along radial direction of the AdS space
down to the horizon and some can extend to the boundary. We define
the number of strings that extend to the boundary to be $k_h$ and
the number of strings extending radially to the horizon to be
$k_r$. The restriction of $k_h$ and $k_r$ is due to the force
condition of the string configuration~(see Ref.~\cite{bch} for
details).

In this article, we consider the holographic model of multiquarks
in the Sakai-Sugimoto (SS) model~\cite{ss lowE, ss more} similar
to the configurations considered in Ref.~\cite{bch}.  The
background metric of the bulk spacetime in the SS model in a
deconfined phase at finite temperature is given by
\begin{equation}
ds^2=\left( \frac{u}{R_{D4}}\right)^{3/2}\left( f(u) dt^2 +
\delta_{ij} dx^{i}
dx^{j}+{dx_4}^2\right)+\left(\frac{R_{D4}}{u}\right)^{3/2}\left(u^2
d\Omega_4^2 + \frac{du^2}{f(u)}\right).\\ \nonumber
\end{equation}
\noindent The four-form field strength, the dilaton, and the
curvature radius of the spacetime are
\begin{equation}
F_{(4)}=\frac{2\pi N}{V_4} {\epsilon}_4, \quad \quad e^{\phi}=g_s
\left( \frac{u}{R_{D4}}\right)^{3/4} ,\quad\quad R_{D4}^3\equiv
\pi g_s N l_{s}^3,\nonumber
\end{equation}
\noindent respectively, where $f(u)\equiv 1-u_{T}^{3}/u^3$,
$u_T=16{\pi}^2 R_{\text{D4}}^3 {T^2} /9$.  $x_4$ is the
compactified coordinate transverse to the probe
D8/$\overline{\text{D8}}$ branes with arbitrary periodicity $2\pi
R$.  The volume of the unit four-sphere $\Omega_4$ is denoted by
$V_4$ and the corresponding volume 4-form by $\epsilon_4$.
$F_{(4)}$ is the 4-form field strength, $l_s$ is the string length
and $g_s$ is the string coupling.

In the SS model, the chiral symmetry dynamics is taken into
account, by construction, in the form of the dynamics of the
flavour branes, D8 and $\overline{\text{D8}}$. The DBI action of
D8-$\overline{\text{D8}}$ is
\begin{eqnarray}
S_{D8} & = & -\mu_{8}\int d^{9}X e^{-\phi}Tr\sqrt{-det(g_{MN}+2\pi
\alpha^{\prime}F_{MN})}
\end{eqnarray}
where $F_{MN}$ is the field strength of the flavour group
$U(N_{f})$ on the branes.  It is given by
\begin{equation}
F = d\mathcal{A} + i \mathcal{A}\wedge \mathcal{A}.
\end{equation}
The $U(N_f)$ gauge field $\mathcal{A}$ can be decomposed into
$SU(N_f)$ part $A$ and $U(1)$ part $\hat{A}$:
\begin{equation}
\mathcal{A}=A+\frac{1}{\sqrt{2N_f}}\hat{A},
\end{equation}
where only the diagonal $U(1)$ will be turned on here.  Lastly,
$g_{MN}$ is the induced metric on the D8-branes world volume.

\begin{figure}
\centering
\includegraphics[width=1.0\textwidth]{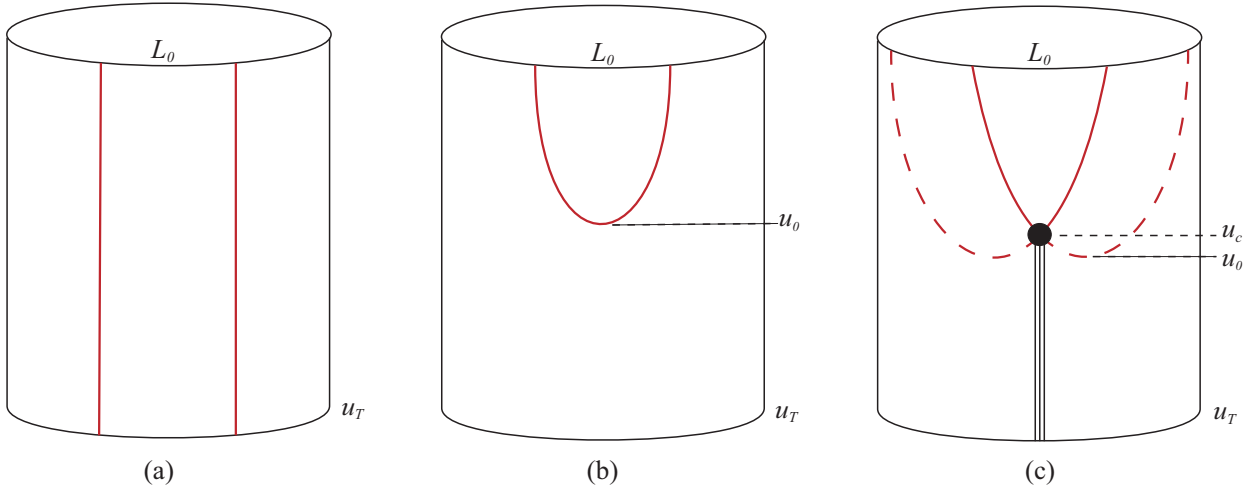} \caption{Different
 configurations of D8 and $\overline{\text{D8}}$-branes in the background
 field following the Sakai-Sugimoto model that are dual to the phases of
 (a) ${\chi}_S$-QGP, (b) vacuum and (c) multiquark phase. } \label{phase}
\end{figure}

In the deconfined phase, the equation of motion from the action of
D8-$\overline{\text{D8}}$ provides 3 possible configurations: (i)
connected D8-$\overline{\text{D8}}$ without sources in the bulk
representing the vacuum state and (ii) the parallel configuration
of both D8-branes and $\overline{\text{D8}}$ representing the
$\chi_S$-QGP.  Another stable configuration~(iii) is the connected
D8-$\overline{\text{D8}}$ branes with the D4-brane as the baryon
vertex submerged and localized in the middle of the D8 and
$\overline{\text{D8}}$.\footnote{Actually, the quark matter,
represented by the connected D8-$\overline{\text{D}8}$ branes with
radial strings stretching out to the horizon, is another possible
configuration satisfying the equation of motion.  However, it was
found that this phase is thermodynamically unstable to density
fluctuations by Bergman, Lifschytz, and Lippert~\cite{bll}.}  In
this model, we assume that the hanging strings shrink to
approximately zero and the only apparent strings are the $k_{r}$
radial strings.  The three configurations are shown in
Fig.~\ref{phase}. We will consider the thermodynamic properties of
only the last multiquark configuration.
The action of the exotic multiquark phase is given by
\begin{equation}
S  = S_{D8}+S_{D4}+\tilde{S}_{F1},\label{total_action}
\end{equation}
where $S_{D8}$ is the DBI action of the connected D8-branes,
$S_{D4}$ represents the DBI action of the D4-brane wrapped on
$S^{4}$ and $\tilde{S}_{F1}$ is the action of $k_r$ radial strings
extending from the baryon vertex down to the horizon. For
simplicity, we ignore the distortion of the baryon vertex due to
the Chern-Simon term~\cite{Callan,Guijosa}.

The DBI action of the D8-$\overline{\text{D8}}$-brane coupled to
the diagonal U(1) gauge field is given by
\begin{eqnarray}
S_{D8}& = & {\mathcal N} \int^{\infty}_{u_{c}} du ~u^{4}
\sqrt{f(u)(x^{\prime}_{4}(u))^{2}+u^{-3}(1-(\hat{a}^{\prime}_{0}(u))^{2})},
\end{eqnarray}
where the constant ${\mathcal N}=(\mu_{8}\tau N_{f}\Omega_{4}V_{3}
R^{5})/g_{s}$, and the rescaled $U(1)$ diagonal field $\hat{a} = 2
\pi \alpha^{\prime}\hat{A}/(R\sqrt{2 N_{f}})$.  Position of the
vertex is denoted by $u_{c}$, it is determined from the
equilibrium condition of the D8-D4-strings configuration~(see
Appendix A of Ref.~\cite{bch}).  The source action of the D4 and
strings, $S_{D4}+\tilde{S}_{F1}$, are given by
\begin{eqnarray}
S_{source}& = & {\mathcal N}d\left[
\frac{1}{3}u_{c}\sqrt{f(u_{c})}+n_{s}(u_{c}-u_{T})\right],
\label{eq:source}
\end{eqnarray}
where $n_{s}$ is the number of radial strings $k_{r}$ in the unit
of $N_{c}$.  The number of radial strings $n_{s}$ represents the
colour charges of a multiquark.  For a fixed number of $k_{h}$,
one of the radial strings can merge with another radial string
from another multiquark and form a colour-binding potential
between the two in a similar way holographic meson is formed
between a quark and an antiquark.

The $U_B (1)$ symmetry corresponds to the $U(1)$-diagonal part of
the global flavor symmetry, $U(N_{f})$, which is provided by the
$N_{f}$ flavor branes. Naturally, the baryon chemical potential,
conjugating to the $U_{B}(1)$ charge, in the gauge theory side can
be identified with the boundary value of the zero component of the
gauge field in the flavor branes, i.e. $A_{0}$, conjugating to the
$U(1)$ ``electric" charge. For convenience, our normalized baryon
chemical potential is~\cite{Tanii}

\begin{equation}
\mu=\hat{a}_{0}(\infty).
\end{equation}

In gauge-gravity duality, we identify the grand canonical
potential density in the gauge theory side in the form of the
D8-branes action evaluated with the classical solution~\cite{Kim}:
\begin{eqnarray}
\Omega(\mu) & = & \frac{1}{{\mathcal
N}}S_{D8}[T,x^{\prime}_{4}(u),\hat{a}_{0}(u)]_{cl}
\end{eqnarray}
With the additional source term, the free energy is in the form of
the combination of the Legendre-transform of the grand potential
and the source action, Eqn.~(\ref{eq:source}). The baryon chemical
potential is simply the derivative of the free energy with respect
to its conjugate, i.e. the baryon number density, at a particular
temperature:
\begin{equation}
\mu =\frac{\partial}{\partial d}\frac{1}{\mathcal N}
\left(\tilde{S}_{\text{D8}}[T,x^{\prime}_4
(u),d(u)]_{cl}+S_{\text{source}}(d,u_c)\right)
\end{equation}
where the Legendre-transformed action $\tilde{S}_{\text{D8}}$ is
given by
\begin{eqnarray}
\tilde{S}_{\text{D8}}[T,x^{\prime}_{4}(u),d(u)] &=
&S_{D8}[T,x^{\prime}_{4}(u),\hat{a}_{0}(u)]+\mathcal{N}
\int_{u_c}^{\infty}
d(u) \hat{a}'_{0} du\\
&=& \mathcal \int_{u_c}^{\infty} du u^{4}
\sqrt{f(u)(x^{\prime}_{4}(u))^2
+u^{-3}}\sqrt{1+\frac{d(u)^2}{u^5}},
\end{eqnarray}
where $d(u)$ is the electric displacement.  It is a constant of
the configuration given by
\begin{equation}\label{d_const}
d(u)=-\frac{1}{\mathcal{N}}\frac{\del S_{D8}}{\del \hat{a}'_{0}
(u)}=\frac{u\hat{a}'_{0} (u)}{\sqrt{f(u)(x'_4
(u))^2+u^{-3}(1-(\hat{a}'_{0}(u))^2)}}=\text{const.}
\end{equation}
Note that the Legendre transformation changes the dependence on
the variable $\hat{a}_{0}(u)$ in $S_{D8}$ to $d(u)$ in
$\tilde{S}_{D8}$.  As a result, the grand potential as a function
of the baryon chemical potential is transformed into the free
energy as a function of the baryon number density.  Another
constant of the configuration is
\begin{eqnarray}\label{x4}
(x^{\prime}_{4}(u))^{2}& = & \frac{1}{u^{3}f(u)}\Big[
\frac{f(u)(u^{8}+u^{3}d^{2})}{F^{2}}-1 \Big]^{-1}=\text{const.},
\label{eq:x4}
\end{eqnarray}
where $F$ is a function of $u_c$, $d$, $T$ and $n_s$, given by
\begin{equation}
F^{2} = u_{c}^{3} f_{c} \left( u_{c}^{5}+d^2 -\frac{d^2
\eta_{c}^{2}}{9 f_{c}}\right),
\end{equation}
where $\eta_{c}\equiv 1+\frac{1}{2}\left(\frac{u_T}{u_c}\right)^3
+3 n_s \sqrt{f_c}$. For convenience, here and henceforth, $f$,
$f_c$ and $f_0$ are used to represent $f(u)$, $f(u_c)$ and
$f(u_0)$, respectively. Note that this form of $F$ is derived from
the force condition at the cusp $u_c$.  The detailed calculations
are given in the Appendix of Ref.~\cite{bch}.

With the equation of motion for $x_4$, Eqn.~(\ref{eq:x4}), and the
separation between D8- and $\overline{\text{D8}}$-branes $L_{0}$
being fixed to $L_{0}=2\int_{uc}^{\infty} x^{\prime}_{4}(u) du
=1$, we obtain~\cite{bll}
\begin{equation}
\mu=\int_{u_c}^{\infty} \hat{a}'_{0}
(u)+\frac{1}{\mathcal{N}}\frac{\partial S_{source}}{\partial
d}\bigg{|}_{T,L_{0},u_{c}},
\end{equation}
where the second term is the contribution from the sources,
$\mu_{source}$.  From these relations, we can then study the
thermodynamic properties of the multiquark phase. The phase
diagram of the multiquark nuclear phase is studied in
Ref.~\cite{bch} when the colour-binding interaction is neglected.
It is found that multiquarks are preferred thermodynamically over
the other gluon-deconfined phases for the large density and
intermediate temperature below the chiral symmetry restoration
temperature.

\section{Calculations of the equation of state}

Thermodynamic properties of the nuclear/exotic matter phase can be
described by the equation of state. First, we will investigate the
relations between the pressure and the number density.  From the
previous section~(see also Ref.~\cite{bch}), the grand potential
density and the chemical potential of the nuclear/exotic matters
are given by
\begin{eqnarray}
\Omega & = & \int^{\infty}_{u_{c}}du {\displaystyle{\left[
1-\frac{F^{2}}{f(u)(u^{8}+u^{3}d^{2})}\right]^{-1/2}}}
\frac{u^{5}}{\sqrt{u^{5}+d^{2}}}, \label{eq:Grand}\\
\mu & = & \int^{\infty}_{u_{c}}du {\displaystyle{\left[
1-\frac{F^{2}}{f(u)(u^{8}+u^{3}d^{2})}\right]^{-1/2}}}\frac{d}{\sqrt{u^{5}+d^{2}}}
+ \frac{1}{3}u_{c}\sqrt{f(u_{c})}+n_{s}(u_{c}-u_{T}) \label{eq:mu}
\end{eqnarray}
\noindent respectively.

Since the differential of the grand potential $G_{\Omega}$ can be
written as
\begin{equation}
dG_{\Omega} = -P dV -S dT-N d\mu
\end{equation}
where the state parameters describing the system $P$, $V$, $S$,
$T$, $N$ are the pressure, volume, entropy, temperature, the total
number of particles of the system respectively.  Since the change
of volume is not our concern, we define the volume density of
$G_{\Omega}$, $S$ and $N$ to be $\Omega$, $s$ and $d$,
respectively. Therefore, we have, at a particular $T$ and $\mu$,
\begin{equation}
P=-G_{\Omega}/V \equiv -\Omega(T,\mu).
\end{equation}
By assuming that the multi-quark states are spatially uniform, we
obtain
\begin{equation}
d=\frac{\partial P}{\partial \mu}(T,\mu).
\end{equation}
Using the chain rule,
\begin{equation}
\frac{\partial P}{\partial d}\Bigg\vert_{T}=\frac{\partial
\mu}{\partial d}\Bigg\vert_{T}\; d,
\end{equation}
so that
\begin{equation}
P(d,T,n_s)=\mu(d,T,n_s)~d -\int_{0}^{d} \mu(d',T,n_s)
~\text{d}(d'),\label{pmud}
\end{equation}
where we have assumed that the regulated pressure is zero when
there is no nuclear matter, i.e. $d=0$.

In the limit of very small $d$, $u_c$ approaches $u_0$, $\eta_c$
becomes $\eta_{0}+\mathcal{O}(d)$, where $\eta_{0}$ is defined to
be $\eta_{c}$ with $u_c$ replaced by $u_0$.  From
Eqn.~(\ref{eq:mu}), the baryon chemical potential can then be
approximated to be
\begin{equation}
\mu-\mu_{source}\simeq d{\Bigg\lbrace \int_{u_c}^{\infty} du\left[
1-\frac{u_{0}^{8} f_{0}}{f u^{8}}-\frac{f_{0} u_{0}^{3}\left(
1-\frac{\eta_{0}^{2}}{9 f_0}-\frac{u_{0}^{5}}{u^{5}}\right)d^2}{f
u^8}\right]^{-1/2} u^{-5/2} \left(1-\frac{d^2}{2u^{5}}\right)
\Bigg\rbrace},
\end{equation}
where
$\mu_{source}=\frac{1}{3}u_{c}\sqrt{f(u_{c})}+n_{s}(u_{c}-u_{T})$,
and we have neglected the higher order terms of $d$.  By using the
binomial expansion, the above equation becomes
\begin{eqnarray}
\mu-\mu_{source} & \simeq & d{\Bigg\lbrace\int_{u_0}^{\infty} du
\frac{u^{-5/2}}{\sqrt{1-\frac{f_{0} u_{0}^8}{f u^{8}}}}\left[ 1+
\left(\frac{f_0
u_{0}^{3}}{fu^8-f_{0}u_{0}^{8}}\left(1-\frac{\eta_{0}^{2}}{9f_0}-\frac{u_{0}^{5}}{u^{5}}\right)-\frac{1}{u^5}\right)\frac{d^2}{2}\right]\Bigg\rbrace}\nonumber\\
& = & \alpha_{0} d - \beta_{0}(n_s) d^3, \label{muofd}
\end{eqnarray}
where
\begin{eqnarray}
\alpha_{0}& \equiv &\int_{u_0}^{\infty} du
\frac{u^{-5/2}}{1-\frac{f_{0}u_{0}^8}{fu^8}} ~, \\
\beta_{0}(n_s)& \equiv &\int_{u_0}^{\infty} du
\frac{u^{-5/2}}{2\sqrt{1-\frac{f_{0}u_{0}^8}{f u^{8}}}}
\left(\frac{f_0
u_{0}^{3}}{fu^8-f_{0}u_{0}^{8}}\left(1-\frac{\eta_{0}^{2}}{9f_0}-\frac{u_{0}^{5}}{u^{5}}\right)+\frac{1}{u^5}\right).
\end{eqnarray}
By substituting Eqn.\eqref{muofd} into Eqn.\eqref{pmud}, we can
determine the pressure in the limit of very small $d$ as
\begin{equation}
P\simeq \frac{\alpha_{0}}{2} d^2 -\frac{3 \beta_{0}(n_s)}{4} d^4.
 \label{eq:Plow}
\end{equation}

In the limit of very large $d$ and relatively small $T$,
\begin{eqnarray}
\mu -\mu_{source}& = &\int_{u_c}^{\infty} du \left[ 1 -\frac{f_c
u_{c}^{3}}{fu^{3}}\left(\frac{u_{c}^{5}+d^2 -\frac{d^2
\eta_{c}^{2}}{9f_c}}{u^5 +d^2}\right)\right]^{-1/2}
\frac{d}{\sqrt{u^5
+d^2}}\\
& \approx & \int_{u_c}^{\infty} du \frac{d}{\sqrt{u^5 +d^2}} +
\frac{1}{2} u_{c}^{3} f_{c} d^2 \left( 1 -
\frac{\eta_{c}^{2}}{9f_c}\right)\int_{u_c}^{\infty} du
\frac{d}{fu^3 (u^5 + d^2)^{3/2}}\\
& \approx & \frac{d^{2/5}}{5} \frac{\Gamma
\left(\frac{1}{5}\right)
\Gamma\left(\frac{3}{10}\right)}{\Gamma\left(\frac{1}{2}\right)}
+\frac{u_{c}^{3} f_{c}}{10}
\left(1-\frac{\eta_{c}^{2}}{9f_c}\right)d^{-4/5} \frac{\Gamma
\left(-\frac{2}{5}\right)\Gamma\left(\frac{19}{10}\right)}{\Gamma\left(\frac{3}{2}\right)}
\label{eq:muhd}
\end{eqnarray}
\noindent where we have used the fact that the lower limit of
integration $u_{c}^5/d^2$ is approximately zero as $d$ is very
large. Again by using Eqn.\eqref{pmud}, we obtain

\begin{equation}
P \simeq \frac{2}{35}\left( \frac{\Gamma\left(\frac{1}{5}\right)
\Gamma\left(\frac{3}{10}\right)}{\Gamma\left(
\frac{1}{2}\right)}\right) d^{7/5}.\label{eq:Phigh}
\end{equation}
And the energy density can then be found via the relation
$d\rho=\mu d(d)$.

Next we consider the entropy of the multiquarks phase.  From the
differential of the free energy,
\begin{equation}
d F_{E}=-PdV-SdT+\mu dN,
\end{equation}
the entropy is given by
\begin{equation}
S=-\frac{\partial F_{E}}{\partial T}.
\end{equation}
The entropy density can then be written as
\begin{equation}
s=-\frac{\partial \mathcal{F}_E}{\partial T},
\end{equation}
where $\mathcal{F}_E$ is the free energy density which relates to
the grand potential density as $\mathcal{F}_E=\Omega+\mu d$. Since
we have the pressure $P=-\Omega$, we can write
\begin{equation}
s=\frac{\partial P}{\partial T}-\left(\frac{\partial \mu}{\partial
T}\right)d.
\end{equation}

For both small $d$ and large $d$, we can see from the formula of
the pressure (see Eqn.\eqref{eq:Plow},\eqref{eq:Phigh}, noting
that $\alpha_{0},\beta_{0}$ is insensitive to temperature) and the
chemical potential (see Eqn.\eqref{muofd},(\ref{eq:muhd})), that
the dominant contribution comes only from $\mu_{source}$, thus
\begin{equation}
s \simeq -\left(\frac{\partial \mu_{source}}{\partial T}\right)d.
\end{equation}
The baryon chemical potential from the D8-branes is insensitive to
the changes of temperature.  This implies that the main
contribution to the entropy density of the multiquark nuclear
phase comes from the source term namely the vertex and strings.

\begin{figure}
\input{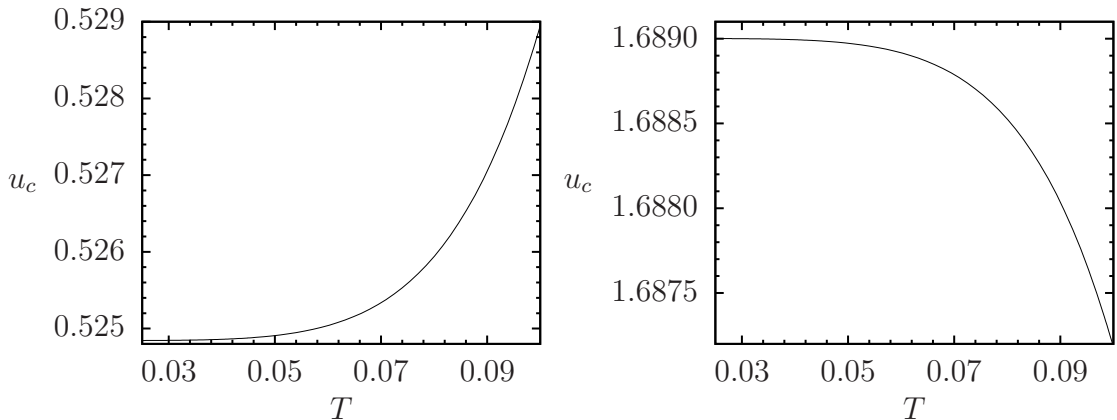}
\caption{The graphs show the relations between $u_c$ and $T$ at
small density (left) and at large density (right).}\label{ucu0}
\end{figure}

Since
\begin{eqnarray}
\frac{\partial \mu_{source}}{\partial T}& = &\frac{\partial
}{\partial T} \left(\frac{1}{3} u_c \sqrt{f(u_c)}+n_s (u_c
-u_T)\right), \\
\frac{\partial \mu_{source}}{\partial T} & \thickapprox &
-\frac{\left(\frac{16\pi^2}{9}\right)^3
T^5}{u_{0}^{2}\sqrt{1-\left(\frac{u_T}{u_0}\right)^{3}}} -n_s
\frac{32\pi^2 T}{9},
\end{eqnarray}
where we have used the fact that $u_{c}$ is approximately constant
with respect to the temperature in the range between the gluon
deconfinement and the chiral symmetry restoration~(see
Fig.~\ref{ucu0}).  Therefore, we obtain
\begin{equation}
s\thickapprox \frac{\left(\frac{16\pi^2}{9}\right)^3 T^5
d}{u_{0}^{2}\sqrt{1-\left(\frac{u_T}{u_0}\right)^{3}}} +n_s
\frac{32\pi^2 T d}{9}.   \label{eq:entropy}
\end{equation}

For small $n_{s}$, the entropy density is proportional to $T^{5}$.
When $n_{s}$ gets larger~(carrying colour charge), the entropy
density becomes dominated by the colour term $s\propto n_{s}T$.
This is confirmed numerically in Section 4.  It has been found
that the entropy density of the $\chi_S$-QGP scales as
$T^{6}$~\cite{bll} corresponding to the fluid of mostly free
quarks and gluons.  We can see that the effect of the colour
charge of the multiquarks as quasi-particles is to make them less
like free particles with the temperature dependence $\sim n_{s}T$,
i.e. much less sensitive to the temperature.

It is interesting to compare the dependence of pressure on the
number density, Eqn.~(\ref{eq:Plow}) and (\ref{eq:Phigh}), to the
confined case at zero temperature studied in Ref.~\cite{ksz}.  The
power-law relations for both small and large density of the
confined and deconfined multiquark phases are in the same
form~(for $n_{s}=0$). The reason is that the main contributions to
the pressure for both phases are given by the D8-branes parts and
they have similar dependence on the density for both phases. For
the deconfined multiquark phase, the additional contributions from
the source terms in Eqn.~(\ref{eq:mu}), $\mu_{source}$, are mostly
constant with respect to the density~(this is because $u_{c}$
becomes approximately independent of $d$ for small and large $d$
limits). Consequently, when we substitute into Eqn.~(\ref{pmud}),
the constant contributions cancel out and affect nothing on the
pressure.

On the contrary, the entropy density for the deconfined phase is
dominated by the contributions from the sources namely the vertex
and strings.  The contribution of the D8-branes is insensitive to
the change of temperature and therefore does not affect the
entropy density significantly.  The additional source terms,
however, depend on the temperature and thus contribute dominantly
to the entropy density.  Once the temperature rises beyond the
gluon-deconfined temperature, entropy density will rise
abruptly~(for sufficiently large density $d$) and become sensitive
to the temperature according to Eqn.~(\ref{eq:entropy}), due to
the release of quarks from colourless confinement appearing as the
sources.  However, we will see later on using the numerical study
in Section 4 that for low densities and for small $n_{s}$, the
numerical value of the entropy density is yet relatively small.

\section{Numerical studies of the thermodynamic relations}

\begin{figure}
\centering
\input{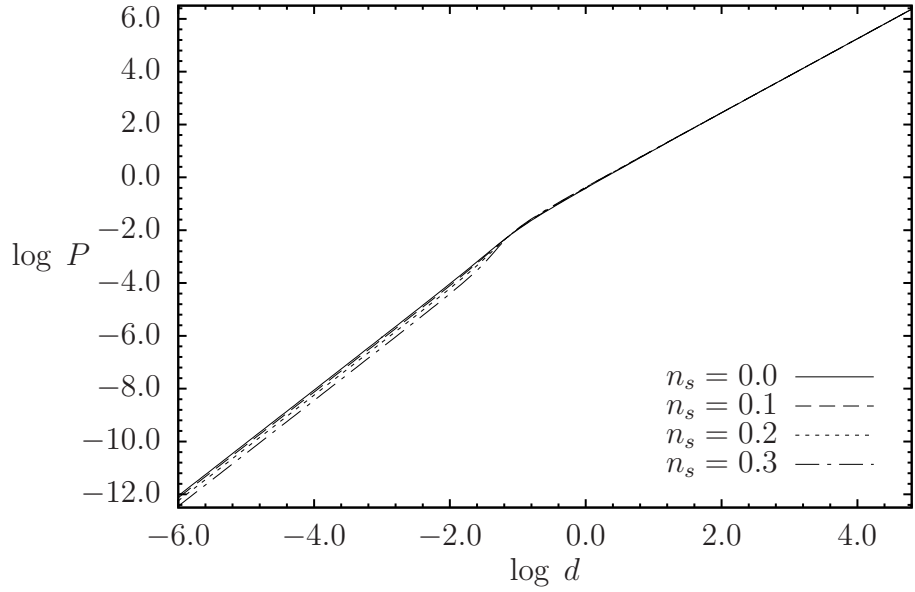}
\caption{Pressure and density in logarithmic scale at
$T=0.03$.}\label{fig1}
\end{figure}

\begin{figure}
\centering
\input{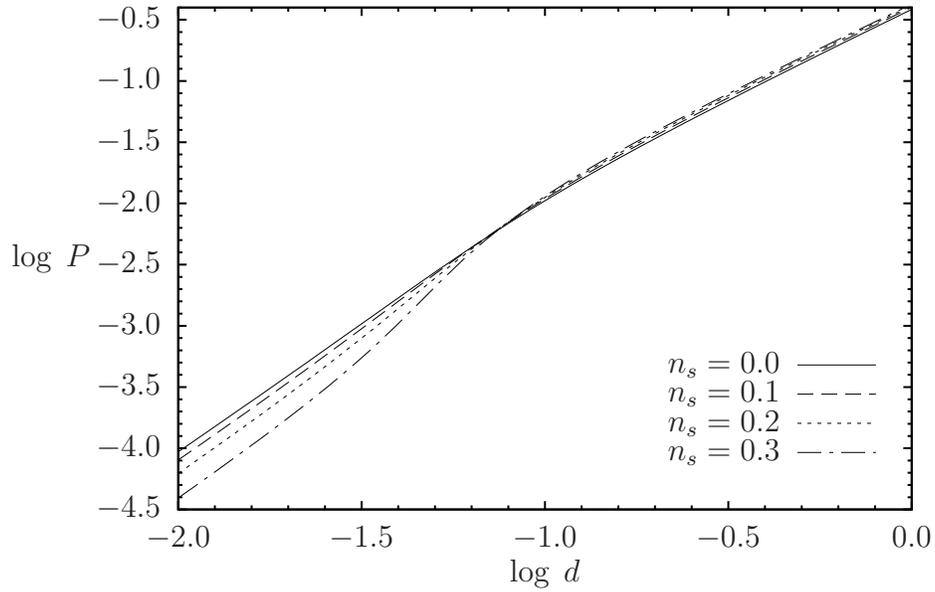}
\caption{Pressure and density in logarithmic scale at $T=0.03$,
zoomed in around the transition region.}\label{fig2}
\end{figure}

\begin{figure}
\centering
\input{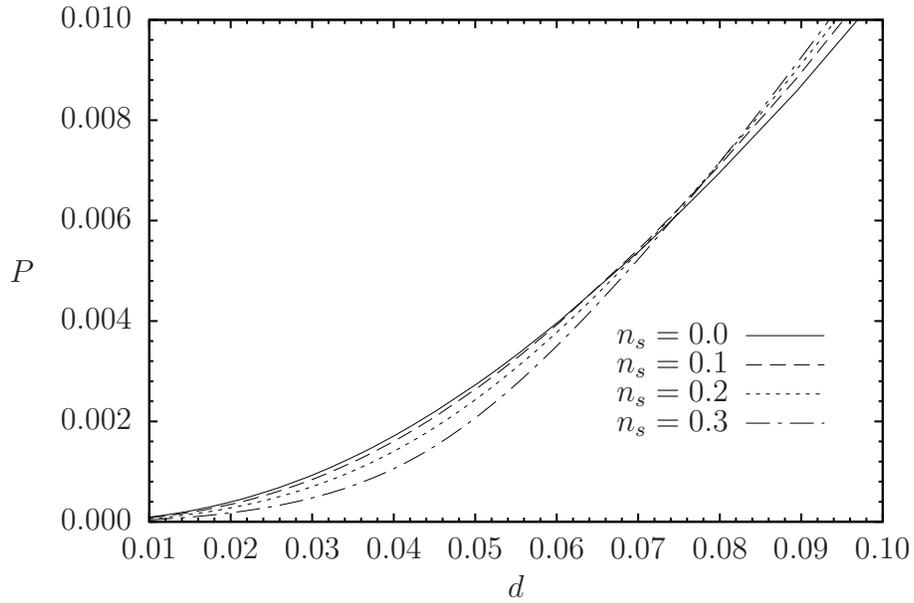}
\caption{Pressure and density in linear scale.}\label{fig5}
\end{figure}

From the analytic approximations in the previous section, we
expect the pressure to appear as straight line in the logarithmic
scale for small and large $d$ with the slope approximately $2$ and
$7/5$ respectively.  The relation between pressure and density of
the multiquarks from the full expressions can be plotted
numerically as are shown in Fig.~\ref{fig1}-\ref{fig5}.  The
pressure does not really depend much on the temperature and we
therefore present only the plots at $T=0.03$.  Remarkably, the
transition from small to large $d$ is clearly visible in the
logarithmic-scale plots. The transition occurs around $d_{c}
\simeq 0.072$. Interestingly, as is shown in Fig.~\ref{fig5}, the
multiquarks with larger $n_{s}$ has lower pressure than the ones
with smaller $n_{s}$ for $d < d_{c}$ and {\it vice versa}. The
dependence on $n_{s}$ remains to be seen for small $d$ as we can
see from Eqn.~(\ref{eq:Plow}). For large $d$, the
$n_{s}$-dependence is highly suppressed as predicted by
Eqn.~(\ref{eq:Phigh}).

\begin{figure}
\input{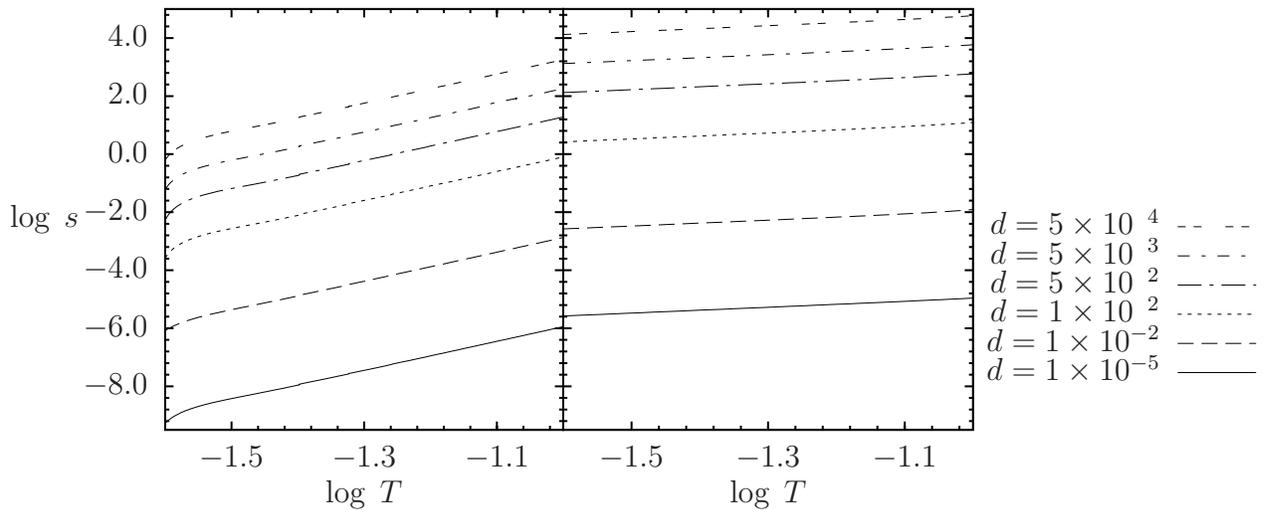}
\caption{Entropy and temperature in logarithmic scale for
$n_{s}=0~(\text{left}), 0.3~(\text{right})$.}\label{fig6}
\end{figure}

The entropy density as a function of the temperature for various
ranges of the density is shown in Fig.~\ref{fig6}.  The
temperature dependence for both small and large $d$ are the same,
$\simeq T^{5}$ at the leading order.  The $d$-dependence is linear
and thus appears as separation of straight lines in the
logarithmic-scale plot.  For $n_{s}>0$, we can see from
Eqn.~(\ref{eq:entropy}) that the linear term in $T$ should become
increasingly important.  This is confirmed numerically as is shown
in Fig.~\ref{fig6}.  The slope of the graph between the entropy
density $s$ and $T$ in the double-log scale for $n_{s}=0$~(the
left plot) and $n_{s}=0.3$~(the right plot) is approximately $5$
and $1$ respectively.  Regardless of the temperature dependence,
it should be noted that the numerical value of the entropy density
for small densities and low $n_{s}$ in Fig.~\ref{fig6} is quite
small.

\begin{figure}
\centering
\input{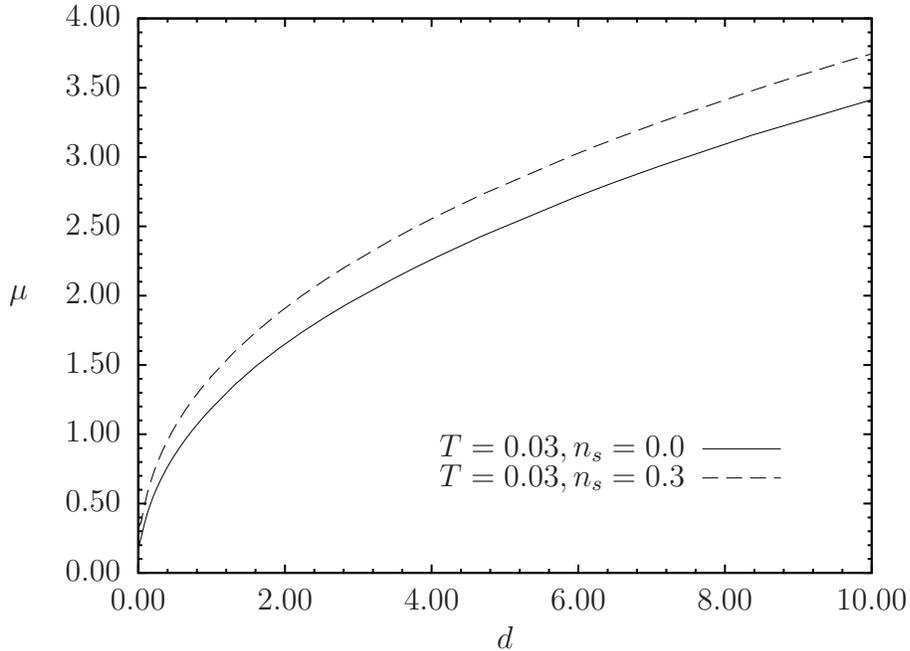}
\caption{The baryon chemical potential and number density in
logarithmic scale at $T=0.03$.}\label{fig6.5}
\end{figure}

Lastly, the relations between the baryon number density and
chemical potential are shown in Fig.~\ref{fig6.5}.  Temperature
has very small effect on these curves and negligible for the range
of temperature between the gluon deconfinement and the
chiral-symmetry restoration.  The baryon chemical potential
depends linearly on the number density for small $d$.  For large
$d$, the relation between the chemical potential and number
density becomes $\mu \approx d^{2/5}$.  Interestingly, the
multiquark quasi-particles behave more like fermions as a result
of being the electric response of the DBI action~\cite{bll}.

\section{Gravitational stability of the dense multiquark star}

When a dying star collapses under its own gravity, it is
generically believed that the degeneracy pressure of either
electrons or neutrons would be able to stop the collapse to form a
white dwarf or a neutron star.  If the star is more massive than
the upper mass limit of the neutron star, it would collapse into a
black hole eventually.  The mass limit of the neutron star is
sensitive to the physics of warm dense nuclear matter but little
is known about the equation of state of nuclear matter under high
temperature and large density. Even though the original mass limit
of the neutron star estimated by Oppenheimer, and Volkoff was only
0.7 solar mass~\cite{tov}, the new limit when the nuclear
interactions are included could be as large as 2.5 solar
mass~\cite{apr}. Under extreme pressure and density, the quarks
within hadrons could be freed and wander around the interior of
the star.  In other words, quarks are effectively deconfined from
the localized hadrons but confined by gravity within the star.
Using the bag model to describe the state of being confined by
gravity but possibly deconfined from the hadrons, it turns out
that quark matter phase, e.g. strange star, is possible under
extreme pressure and density.

However, physics of the deconfinement is largely unknown due to
the non-perturbative nature of the strong interaction and the
difficulty of lattice approach to deal with finite baryon density
situation. The bag model are not always served as a reliable
theoretical tool to explore the behaviour of quarks in the dense
star when the deconfinement exists.  It is therefore interesting
to use the equation of state of the deconfined nuclear matter from
the holographic model to investigate the behaviour of the dense
star as a complementary tool to the bag model and other
approaches.

In this section, we will consider a hypothetical multiquark star
containing only the multiquark matter with uniform constant
temperature.  The relations between pressure and density will be
adopted directly from the holographic model as the equations of
state of the quasi-particles.  Since the pressure and density have
very small temperature dependence for the range of temperatures
under consideration, the results are valid generically.

A study into the gravitational stability of a spherically
symmetric dense star can be performed using the
Tolman-Oppenheimer-Volkoff equation~\cite{tov}. It is known that
the spherically symmetric dense star has metric in a form
\begin{equation}
ds^{2}=A(r)dt^{2}-B(r)dr^{2}-r^{2}d\Omega_{2}.
\end{equation}

After substituting into the Einstein field equation, we obtain the
following relations,
\begin{eqnarray}
B(r)& = & \left(1-\frac{A^{*}(r)}{r}\right)^{-1}, \\
\frac{dA^{*}(r)}{dr}& = & 8\pi\rho r^{2},
\end{eqnarray}
and
\begin{equation}
\frac{dP(r)}{dr} =
-\frac{\left(\rho+P\right)}{2}\frac{A^{\prime}(r)}{A(r)}=
-\frac{\left(\rho+P\right)}{2}\frac{8\pi P
r^{3}+A^{*}(r)}{r(r-A^{*}(r))}.
\end{equation}
The last equation is known as the Tolman-Oppenheimer-Volkoff~(TOV)
equation.  The accumulated mass of the star, $M(r)$, is given by
$A^{*}(r)=2M(r)$.  It has been shown in Ref.~\cite{holo_n_star}
that the chemical potential can be defined through the background
metric in the form of $\mu(r) = \frac{\epsilon_{F}}{\sqrt{A(r)}}$.
It will automatically solve the TOV equation.  Note that the
constant $\epsilon_{F}$ is arbitrary. Since
\begin{equation}
\frac{\mu^{\prime}(r)}{\mu(r)}=-\frac{1}{2}\frac{A^{\prime}(r)}{A(r)},
\end{equation}
the TOV equation becomes
\begin{equation}
\frac{dP(r)}{dr}=\left(\rho+P\right)\frac{\mu^{\prime}(r)}{\mu(r)}.
\end{equation}
Together with the first law of thermodynamics $\rho+P = \mu d$,
the TOV equation then takes the following form,
\begin{equation}
d\mu=\frac{1}{d}\left(\frac{\partial P}{\partial d}\right)d(d).
\end{equation}
Obviously, the chemical potential can be determined, as a function
of the number density:
\begin{equation}
\mu(d)=\int^{d}_{0}\frac{1}{\eta}\left(\frac{\partial P}{\partial
\eta}\right)d\eta+\mu_{onset},   \label{eq:mc}
\end{equation}
where $\mu_{onset}\equiv \mu(d=0)$.  Additionally, considering
from the TOV equation together with the first law of
thermodynamics, the density $d\rho = \mu d(d)$ can be integrated
to
\begin{equation}
\rho(d)=\int^{d}_{0}\left[\int^{\eta}_{0}\frac{1}{\eta^{\prime}}\left(\frac{\partial
P}{\partial
\eta^{\prime}}\right)d\eta^{\prime}+\mu_{onset}\right]d\eta.
\label{eq:rc}
\end{equation}
For a power-law equation of state, $P = k d^{\lambda}$, the
chemical potential, Eqn.~(\ref{eq:mc}), becomes
\begin{equation}
\mu(d)=\frac{\lambda k}{\lambda -1}d^{\lambda-1}+\mu_{onset},
\end{equation}
and eventually the equation of state is given by
\begin{equation}
\rho=\frac{1}{\lambda
-1}P+\mu_{onset}\left(\frac{P}{k}\right)^{1/\lambda}.
\label{eq:rho}
\end{equation}

In our holographic model of multiquarks, the relation between
pressure and density has a unique power-law behaviour, as is also
found in Ref.~{\cite{bll}} for the case of normal
baryon~($n_{s}=0$).  This is shown in Fig.~\ref{fig1}-\ref{fig2}.
For small $d$, $P \propto d^{2}~(n_{s}=0)$ and for large $d$, $P
\propto d^{7/5}$.  The dependence on $n_{s}$ becomes significant
when the density $d$ is small and the equation of state can be
approximated by $P \simeq \alpha d^{2} + \beta d^{4}$.  Since
there are two power-laws governing, we need to match the solutions
from the two regions together~(i.e.~{\it core} and {\it crust}).
The number density where the equation of state changes from the
large-$d$ to the small-$d$ is denoted by $d_{c}$.

\underline{For $n_{s}=0$}, at the transition point $d=d_{c}$, the
energy density is given by Eqn.~(\ref{eq:rho}),
\begin{eqnarray}
\rho_{c} & = &
\frac{k^{\prime}d_{c}^{\lambda^{\prime}}}{\lambda^{\prime}-1} +
\mu_{onset}d_{c},
\end{eqnarray}
where $P=k^{\prime}d^{\lambda^{\prime}}$~(Eqn.~(\ref{eq:Plow})
suggests that $\lambda^{\prime}=2$) is the equation of state of
the small $d$ region.  We recalculate the relation
Eqn.~(\ref{eq:mc}), (\ref{eq:rc}) for the large $d$ region which
match with this $\rho_{c}$ to be
\begin{eqnarray}
\mu & = & \mu_{c} + \lambda k \left( \frac{d^{\lambda -1}}{\lambda
-1}- \frac{d_{c}^{\lambda -1}}{\lambda -1} \right), \label{eq:mc1}  \\
\rho & = & \rho_{c} + \frac{1}{\lambda -1}P+\mu_{c}\left[
\left(\frac{P}{k}\right)^{1/\lambda}-d_{c} \right]+k
d_{c}^{\lambda}-\frac{\lambda k}{\lambda -1}d_{c}^{\lambda
-1}\left(\frac{P}{k}\right)^{1/\lambda}.  \label{eq:rc1}
\end{eqnarray}
Numerical results and Eqn.~(\ref{eq:Phigh}) suggest that $\lambda
= 7/5$ for the large $d$ region.

\underline{For $n_{s}>0$}, assume the equation of state for small
$d$ is in the form of $P=a d^{\lambda_{1}}+b
d^{\lambda_{2}}$~(Eqn.~(\ref{eq:Plow}) suggests that
$\lambda_{1,2}=2,4$), the chemical potential and energy density
for the small $d$ region become
\begin{eqnarray}
\mu& = & \mu_{onset}+\frac{\lambda_{1} a
d^{\lambda_{1}-1}}{\lambda_{1}-1}+\frac{\lambda_{2} b
d^{\lambda_{2}-1}}{\lambda_{2}-1},
\end{eqnarray}
\begin{eqnarray}
\rho& = & \mu_{onset}d+\frac{a
d^{\lambda_{1}}}{\lambda_{1}-1}+\frac{b
d^{\lambda_{2}}}{\lambda_{2}-1}.
\end{eqnarray}
We obtain the transition density in the similar fashion,
\begin{eqnarray}
\rho_{c}& = & \mu_{onset}d_{c}+\frac{a
d_{c}^{\lambda_{1}}}{\lambda_{1}-1}+\frac{b
d_{c}^{\lambda_{2}}}{\lambda_{2}-1}.
\end{eqnarray}

Numerical results show that for large $d$, the effect of $n_{s}$
is negligible.  Therefore, the baryon chemical potential and the
density for the large $d$ region are again given by
Eqn.~(\ref{eq:mc1}) and (\ref{eq:rc1}).  The equations of state,
Eqn.~(\ref{eq:rho}),(\ref{eq:rc1}) as well as the corresponding
relations for $n_{s}>0$ case, are in the mixed form containing
both the quasi-particle nonlinear terms and the linear term.  The
linear term is roughly $\rho_{linear}\approx 2.5P$ and the
quasi-particle term is approximately $\rho_{quasi}\approx
P^{5/7}$.

We can solve the TOV equation when the equations of state are
given as above by starting from the core of the star out to the
surface.  As we go from the center towards the surface of the
star, the density decreases until it reaches a critical value
$\rho_{c}$.  This density corresponds to the number density
$d_{c}$ where the power-law changes from $P\simeq d^{7/5}$ to $P
\simeq d^{2}$~(see Fig.~\ref{fig1}-\ref{fig2}).  For the crust
region where the density $\rho < \rho_{c}$, multiquarks obey a
different equation of state given by Eqn.~(\ref{eq:rho}). The
radius of the core is defined to be the distance $R_{core}$ where
$\rho(R_{core})=\rho_{c}$ and the surface of the star is defined
to be the radial distance $R$ where $\rho(R)=0$.

\begin{figure}[htp]
\centering
\includegraphics[width=0.7\textwidth]{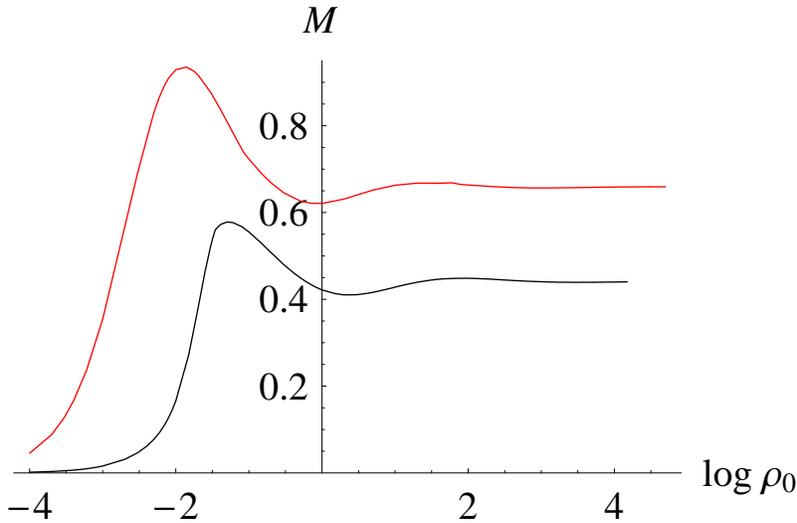}
\caption[masslimit0]{The relation between mass and central density
of the multiquark star for multiquarks with
$n_{s}=0~(\text{upper}), 0.3~(\text{lower})$.}\label{fig7}
\end{figure}

\begin{figure}[htp]
\centering
\includegraphics[width=0.6\textwidth]{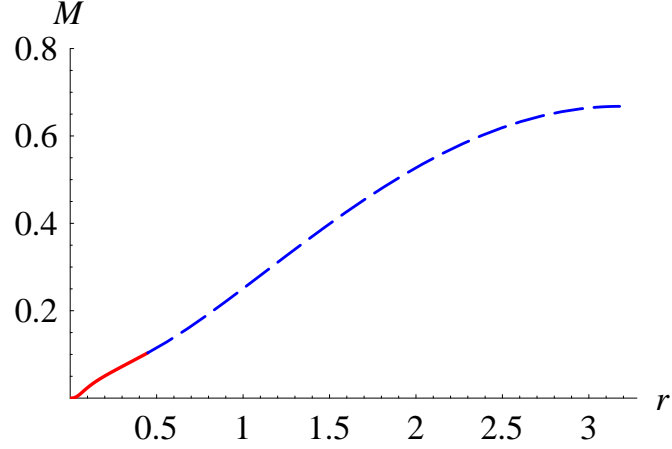}
\caption[star profile]{The accumulated mass distribution in the
hypothetical multiquark star for the central density $\rho_{0}=20$
and $n_{s}=0$.  The inner~(outer) red~(dashed-blue) line
represents the core~(crust) region.} \label{fig8}
\end{figure}

\begin{figure}[htp]
\centering
\includegraphics[width=0.45\textwidth]{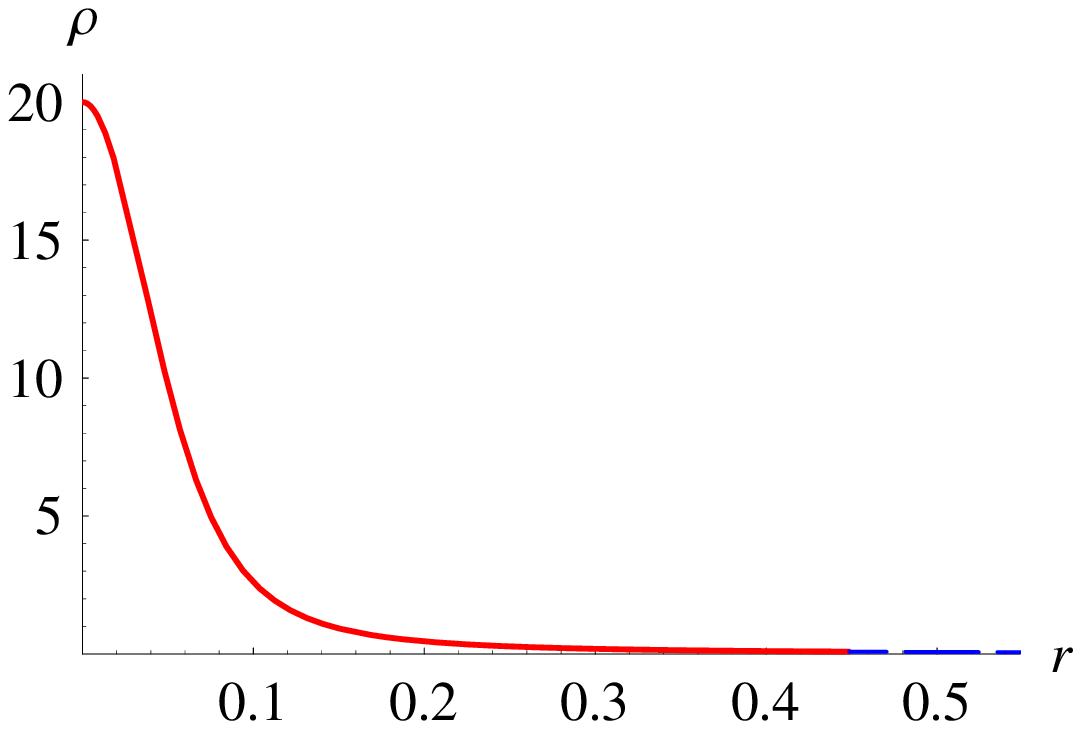} \hfill
\includegraphics[width=0.45\textwidth]{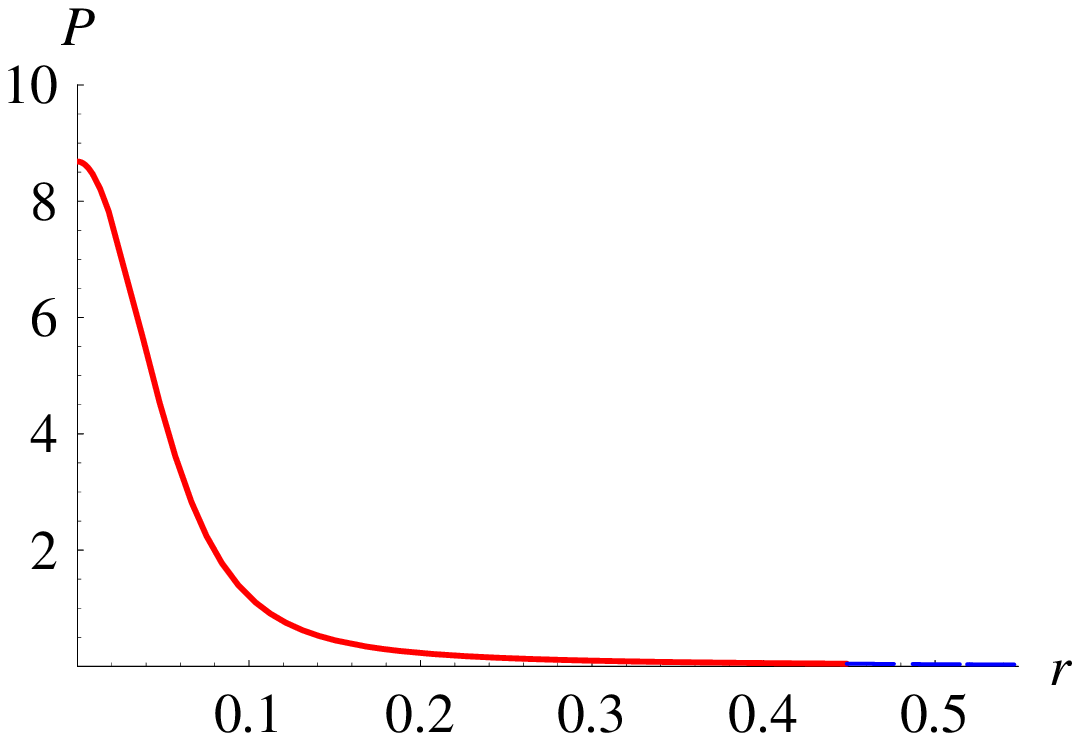}\\
\caption[star profile]{The density, and pressure distribution in
the hypothetical multiquark star for the central density
$\rho_{0}=20$ and $n_{s}=0$.  The inner~(outer) red~(dashed-blue)
line represents the core~(crust) region.} \label{fig8.1}
\end{figure}

\begin{figure}[htp]
\centering
\includegraphics[width=0.45\textwidth]{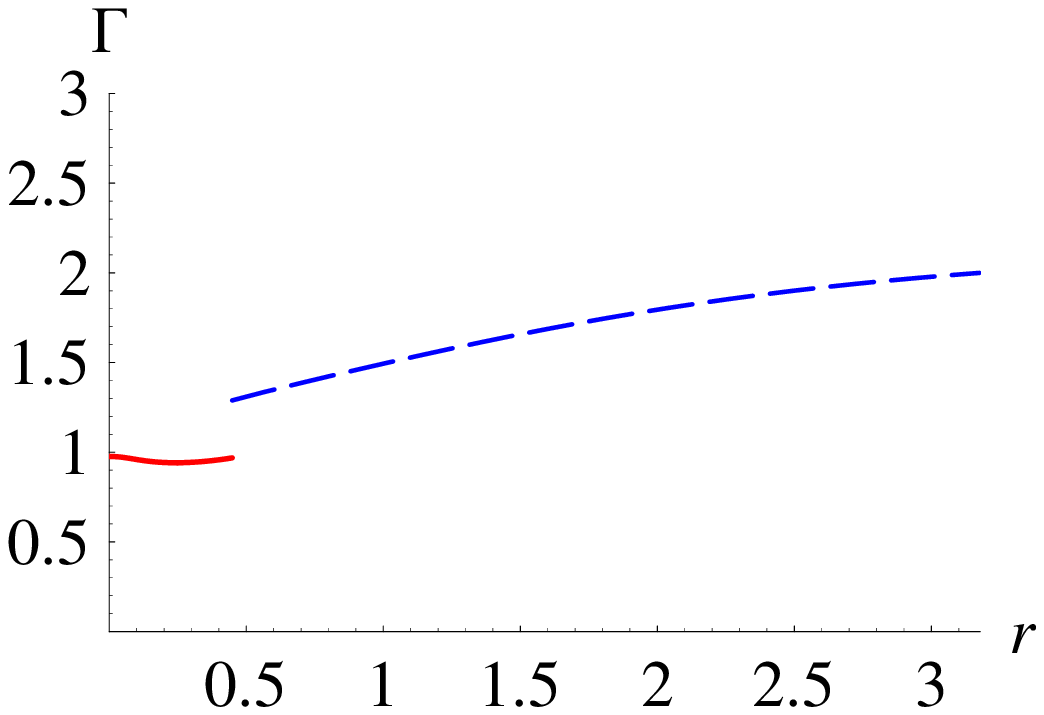} \hfill
\includegraphics[width=0.45\textwidth]{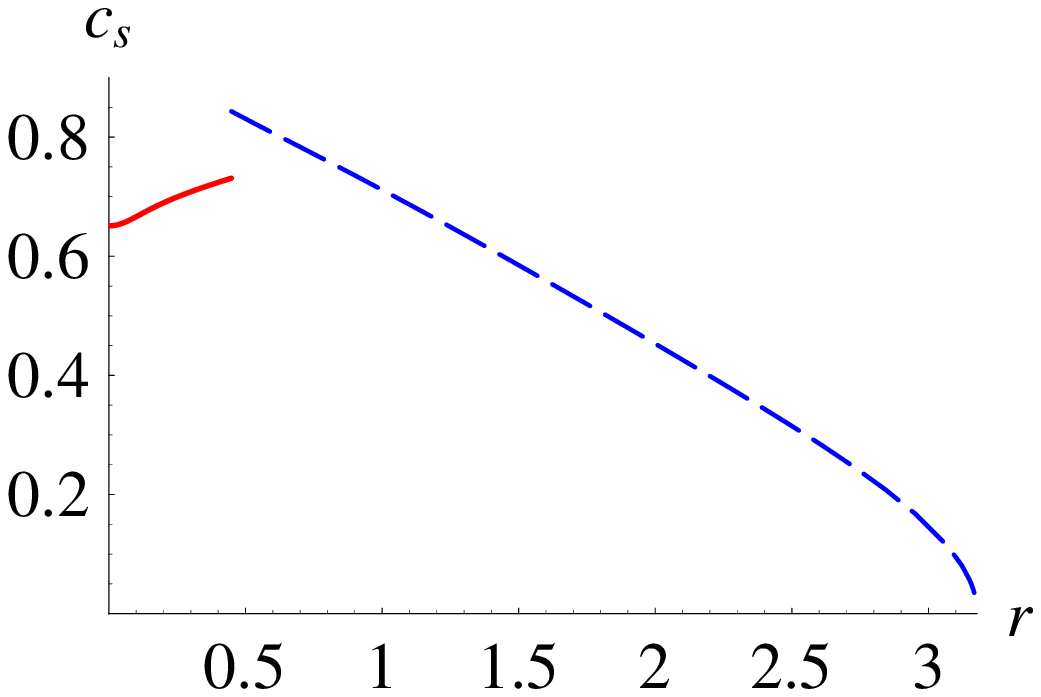}\\
\caption[star profile]{The adiabatic index at constant
entropy~($\Gamma$) and the sound speed~($c_{s}$) distribution in
the hypothetical multiquark star for the central density
$\rho_{0}=20$ and $n_{s}=0$.  The inner~(outer) red~(dashed-blue)
line represents the core~(crust) region.} \label{fig8.5}
\end{figure}

\begin{figure}[htp]
\centering
\includegraphics[width=0.6\textwidth]{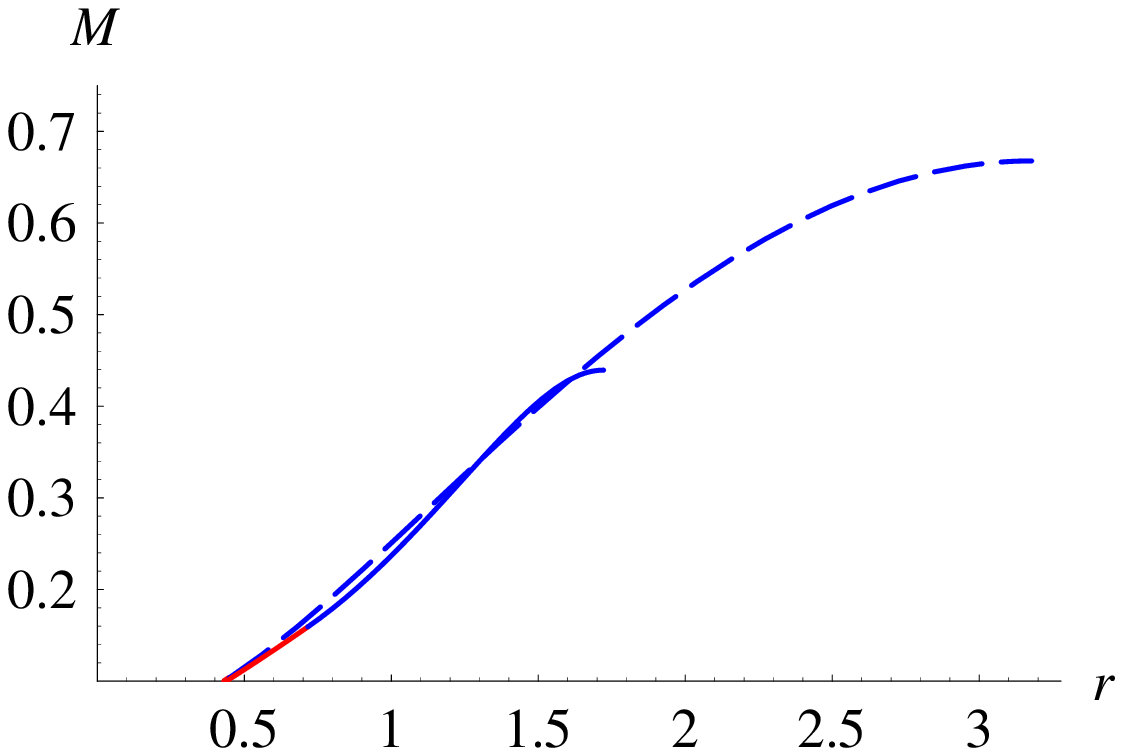}
\caption[star profile]{Comparison of the accumulated mass
distribution in the hypothetical multiquark star for the central
density $\rho_{0}=20$ between $n_{s}=0$ and $0.3$.  The (dashed)
blue line represents the crust region of multiquark star with
$n_{s}=0.3~(0)$.  The red lines represent the core region of which
both cases are almost the same.} \label{fig9}
\end{figure}

\begin{figure}[htp]
\centering
\includegraphics[width=0.45\textwidth]{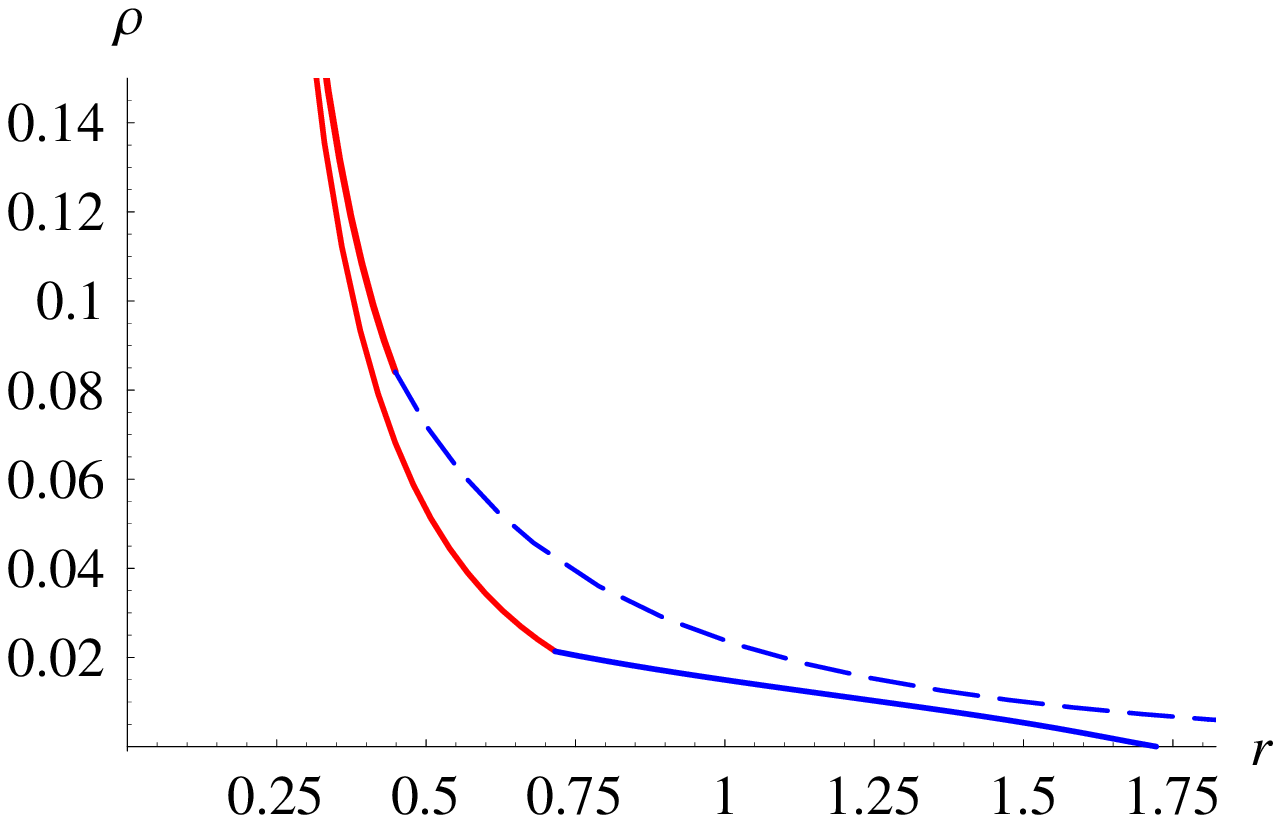} \hfill
\includegraphics[width=0.45\textwidth]{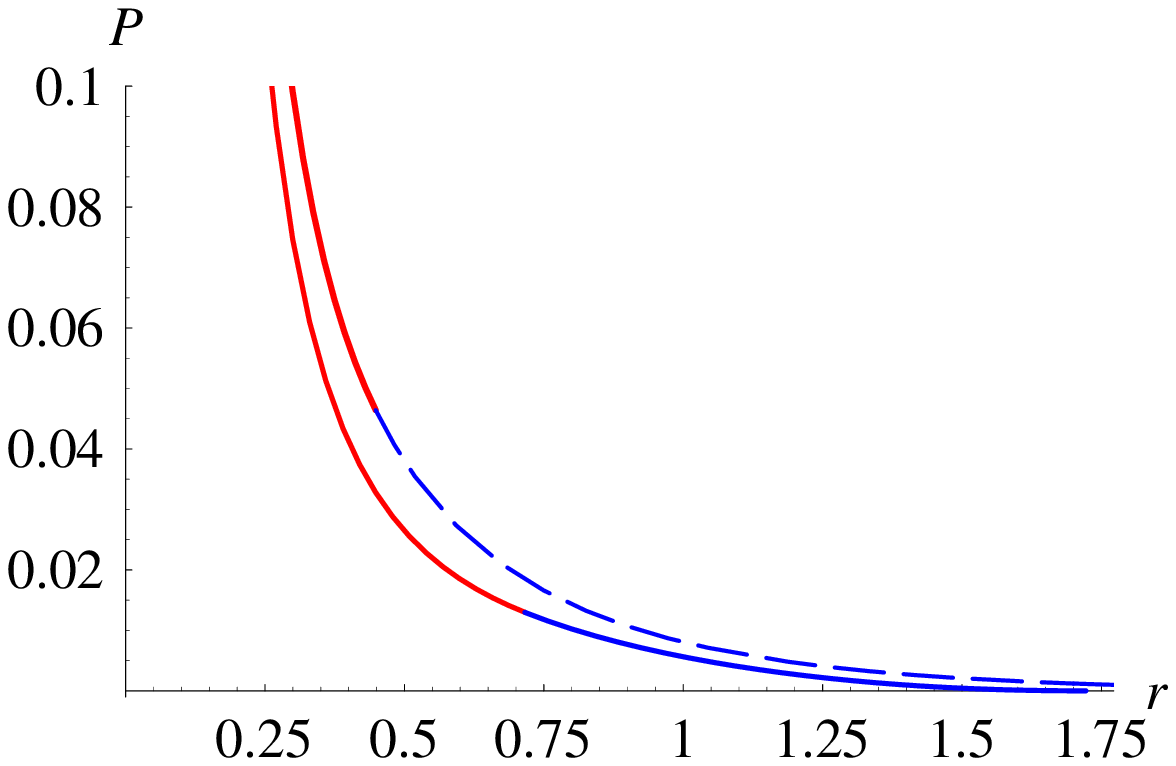}\\
\caption[star profile]{Comparison of the density, and pressure
distribution in the hypothetical multiquark star for the central
density $\rho_{0}=20$ between $n_{s}=0$ and $0.3$.  The (dashed)
blue line represents the crust region of multiquark star with
$n_{s}=0.3~(0)$.  The red lines represent the core region of which
both cases are almost the same.} \label{fig9.1}
\end{figure}

\begin{figure}[htp]
\centering
\includegraphics[width=0.45\textwidth]{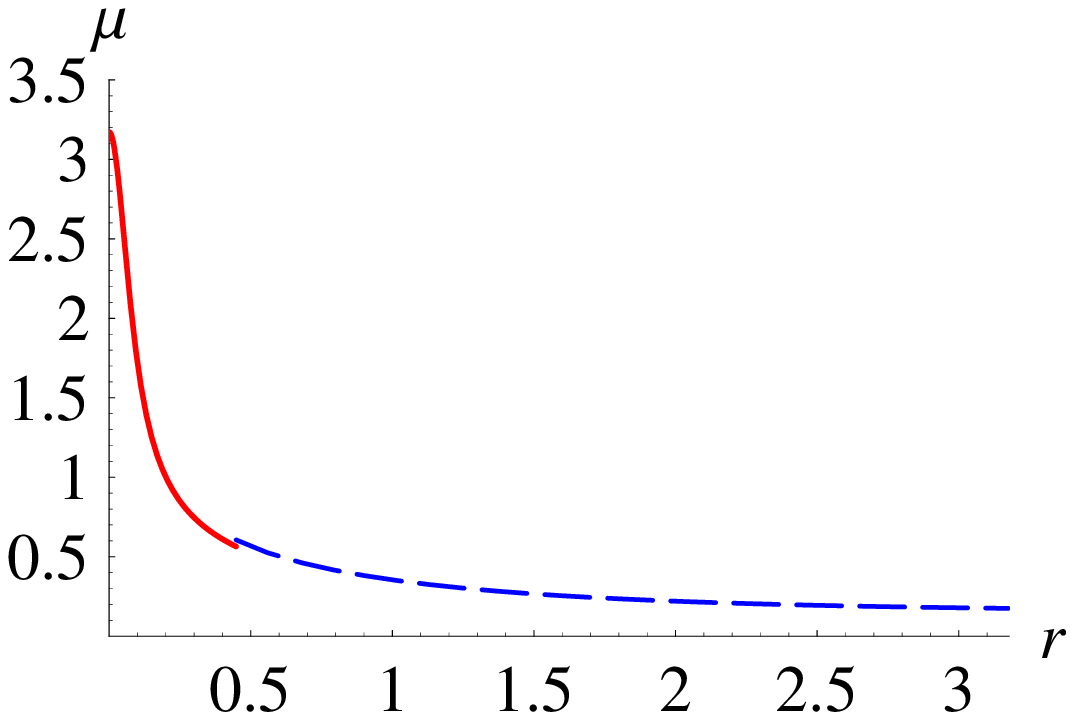} \hfill
\includegraphics[width=0.45\textwidth]{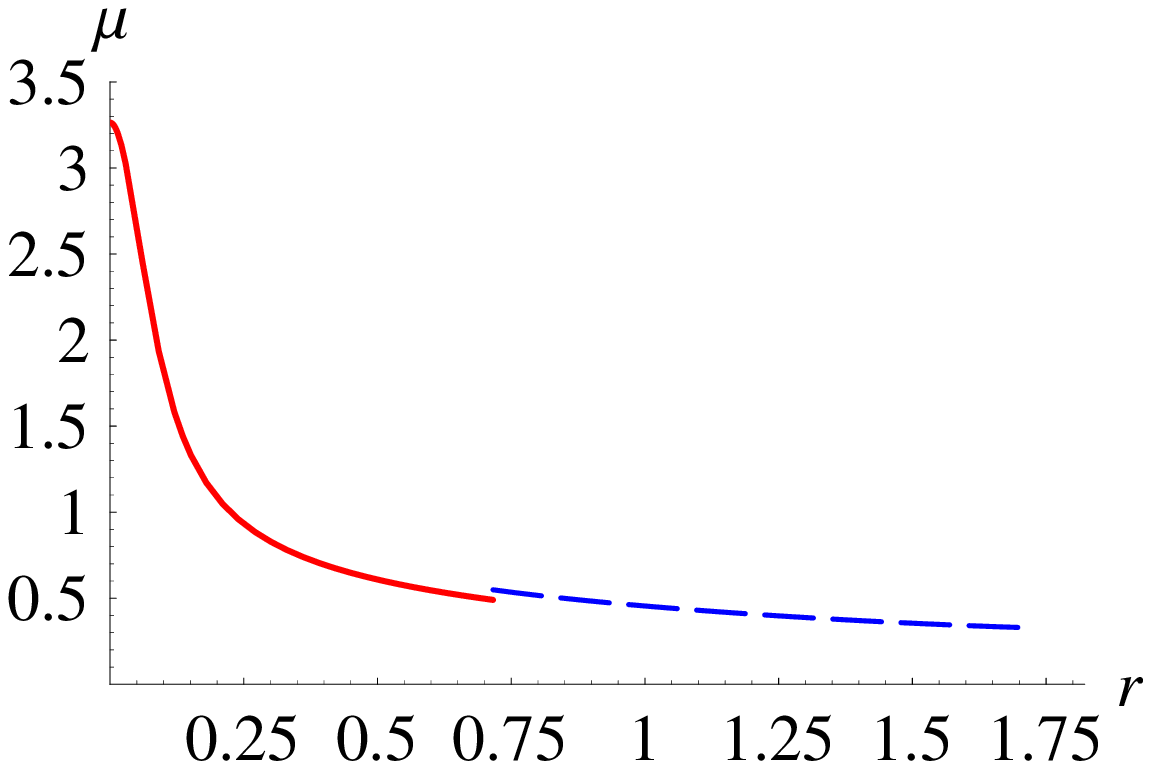}\\
\caption[star profile]{Comparison of the baryon chemical
distributions in the hypothetical multiquark star for the central
density $\rho_{0}=20$ between $n_{s}=0$~(left) and $0.3$~(right).
The solid~(dashed) red~(blue) line represents the core~(crust)
region. } \label{fig9.2}
\end{figure}

\begin{figure}[htp]
\centering
\includegraphics[width=0.45\textwidth]{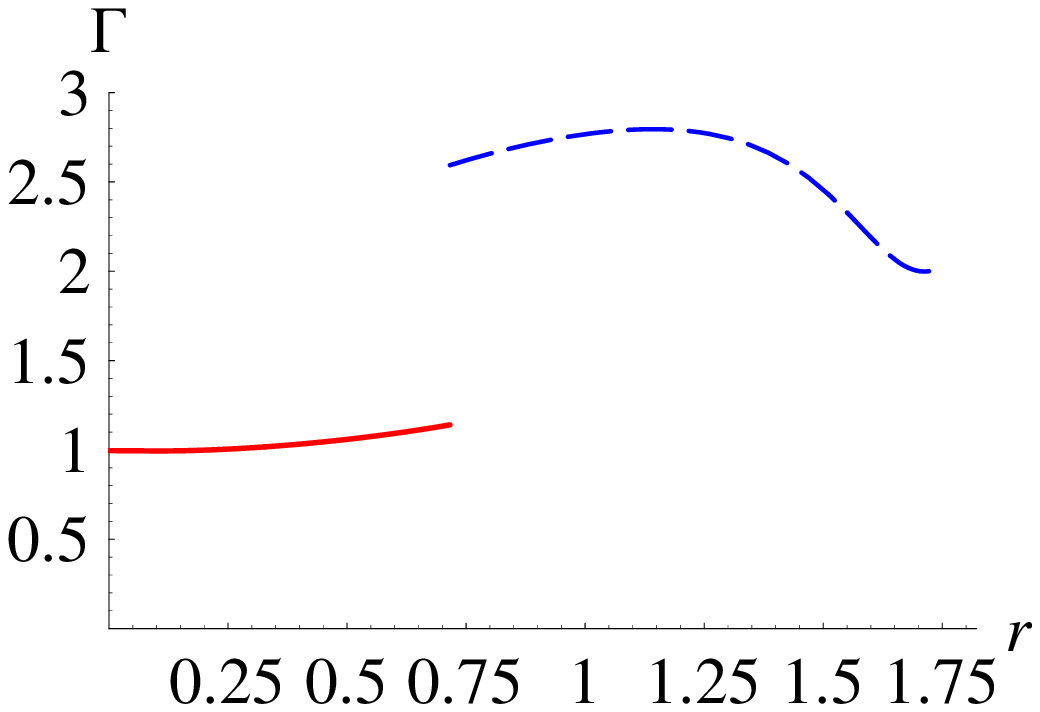} \hfill
\includegraphics[width=0.45\textwidth]{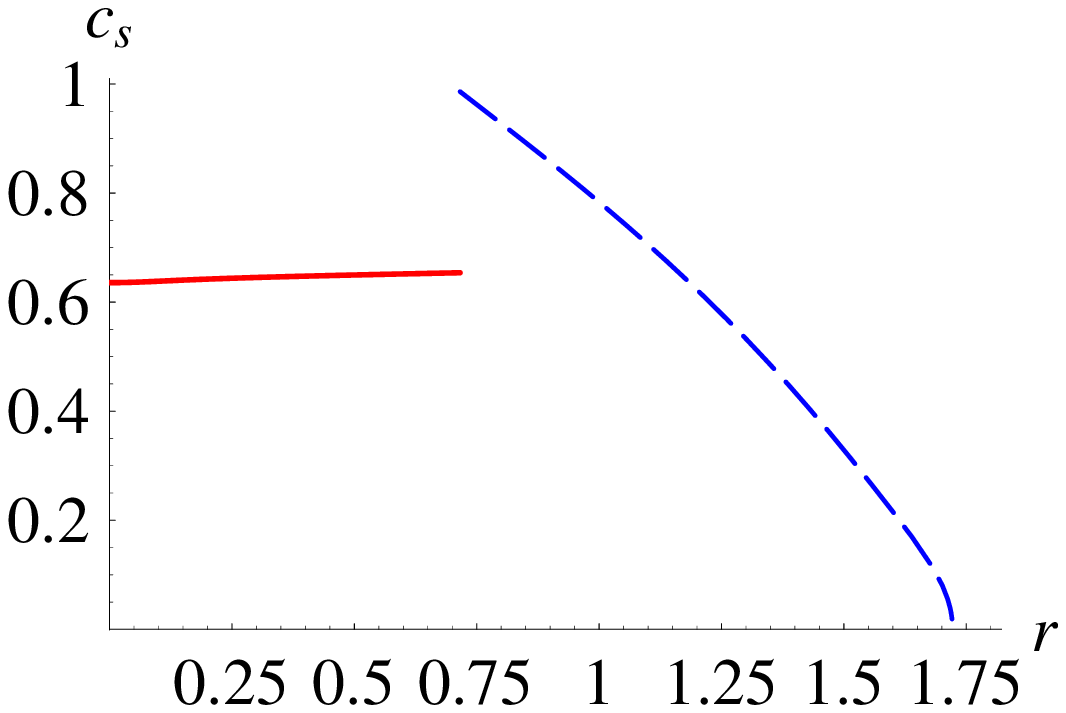}\\
\caption[star profile]{The adiabatic index at constant
entropy~($\Gamma$) and the sound speed~($c_{s}$) distribution in
the hypothetical multiquark star for the central density
$\rho_{0}=20$ and $n_{s}=0.3$.  The inner~(outer)
red~(dashed-blue) line represents the core~(crust) region.}
\label{fig9.5}
\end{figure}

For $n_{s}=0$, numerical fittings suggest that $k=10^{-0.4},
\lambda = 7/5, d_{c}=0.215443, \mu_{c}=0.564374$~(core) and
$k^{\prime}=1, \lambda^{\prime}=2, \mu_{onset}=0.17495$~(crust).
For $n_{s}=0.3$, good fit parameters are $k=10^{-0.4}, \lambda =
7/5, d_{c}=0.086666, \mu_{c}=0.490069$~(core) and $a,b=0.375,
180.0,; \lambda_{1,2}=2,4; \mu_{onset}=0.32767$~(crust). Varying
the central density $\rho_{0}$ of the star, we obtain the
mass-density relation in Fig.~\ref{fig7}.  Each curve has two
maxima, a larger one in the small density region and a smaller one
in the large density region.  Each maximum corresponds to each
power-law of the equation of state, the low density to the crust
and the large density to the core. Interestingly, the contribution
to the total mass of the multiquark star comes dominantly from the
crust.  This is shown in Fig.~\ref{fig8}.  Even though the density
is much lower, the volume of the crust is proportional to the
second power of the radius and thus makes the contribution of the
crust to the total mass larger than the core's.

Figure~\ref{fig8.1} shows the pressure and density distribution
within the multiquark star for the case of $n_{s}=0$ for the
central density $\rho_{0}=20$.  Even though the density and
pressure decrease rapidly with respect to the radius of the star,
they never quite reach zero.  It turns out that when the density
and pressure reach the critical values where the equation of state
changes into the different power-law for small $d$, the crust
region continues for a large fraction of the total radius of the
star.  This makes the crust mass contribution to the total mass of
the star dominant as is shown in Fig.~\ref{fig8}.

Some remarks should be made regarding the hydrodynamic properties
of the multiquark phase~(taken as nuclear liquid).  At constant
temperature and entropy, we can define the adiabatic index
\begin{eqnarray}
\Gamma & \equiv & \frac{\rho}{P}\frac{\partial P}{\partial \rho},
\\
& = & \frac{\rho}{P}c^{2}_{s}
\end{eqnarray}
where $c_{s}$ is the sound speed in the multiquark liquid.  They
depend on the equation of state of the multiquark and their
distributions within the multiquark star are shown in
Fig.~\ref{fig8.5} for $n_{s}=0$.  The sound speed never exceeds
the speed of light in vacuum.  It is also found that the adiabatic
index and the sound speed change within a small fraction as the
central densities are varied for a given $n_{s}$.

The multiquark star with $n_{s}=0.3$~(having colour charges)
converge to a smaller mass and radius at high central
density~(Fig.~\ref{fig9}). Multiquarks with colour charges has
lower pressure~(and therefore smaller density) than the colourless
ones for small density~(Fig.~\ref{fig9.1}). This smaller pressure
makes the coloured multiquark star smaller and thus less massive
than the colourless one.  In more realistic situations, all of the
possible multiquarks with varying $n_{s}$ coexist in the
multiquark phase. The mass limit and mass radius relation will
vary between the two typical cases we consider here. Since the
equations of state are found NOT to be sensitive to the
temperature within the range between the gluon deconfinement and
the chiral symmetry restoration, our results should also be valid
even when the temperature varies within the star~(but not too high
and too low, of course).

\begin{figure}[htp]
\centering
\includegraphics[width=0.45\textwidth]{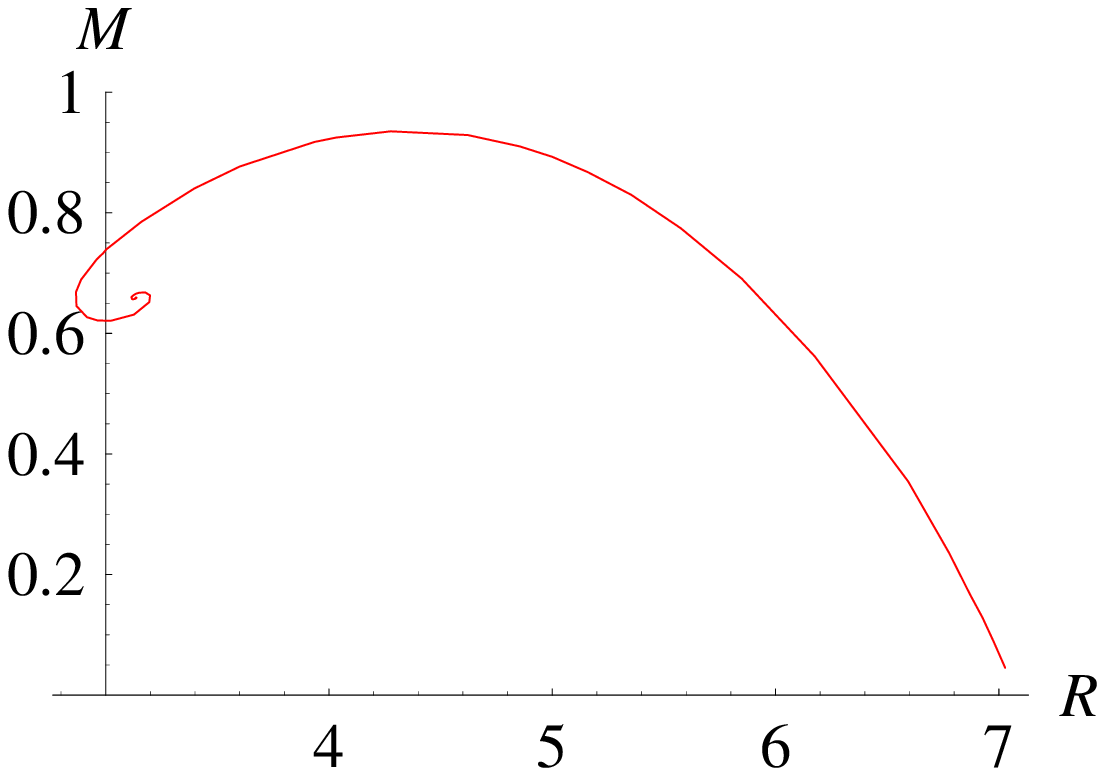} \hfill
\includegraphics[width=0.45\textwidth]{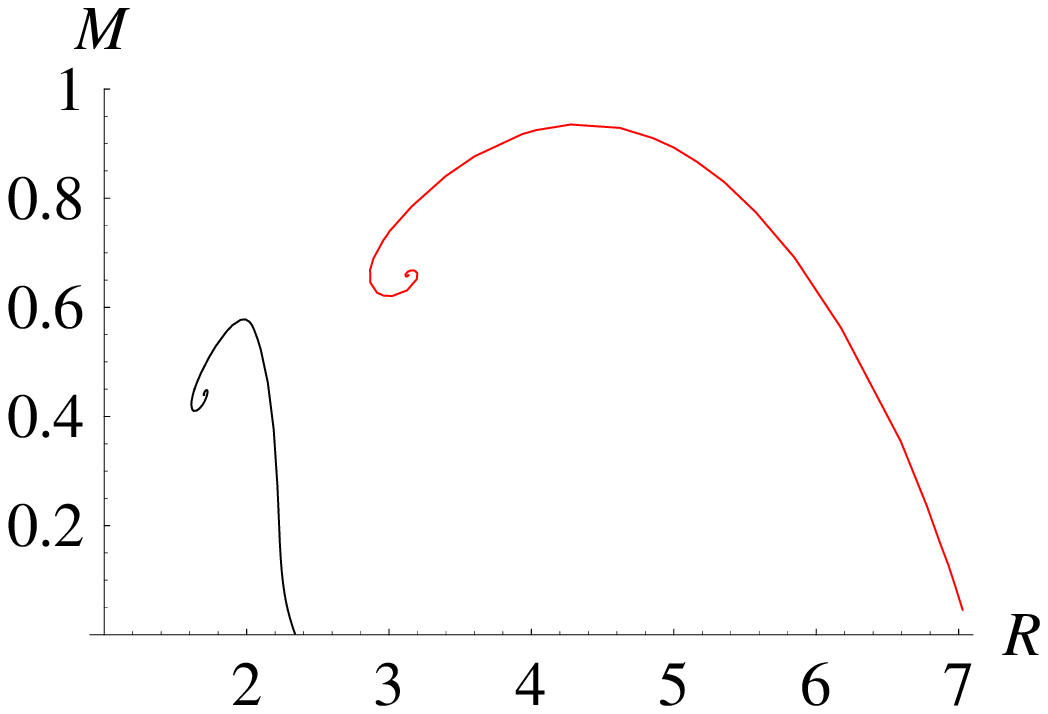} \\
\caption[mass and radius]{The relation between mass and radius of
the multiquark star with (a) $n_{s}=0$, (b) $n_{s}=0$~(red) and
$n_{s}=0.3$~(black). } \label{fig10}
\end{figure}

\begin{figure}[htp]
\centering
\includegraphics[width=0.45\textwidth]{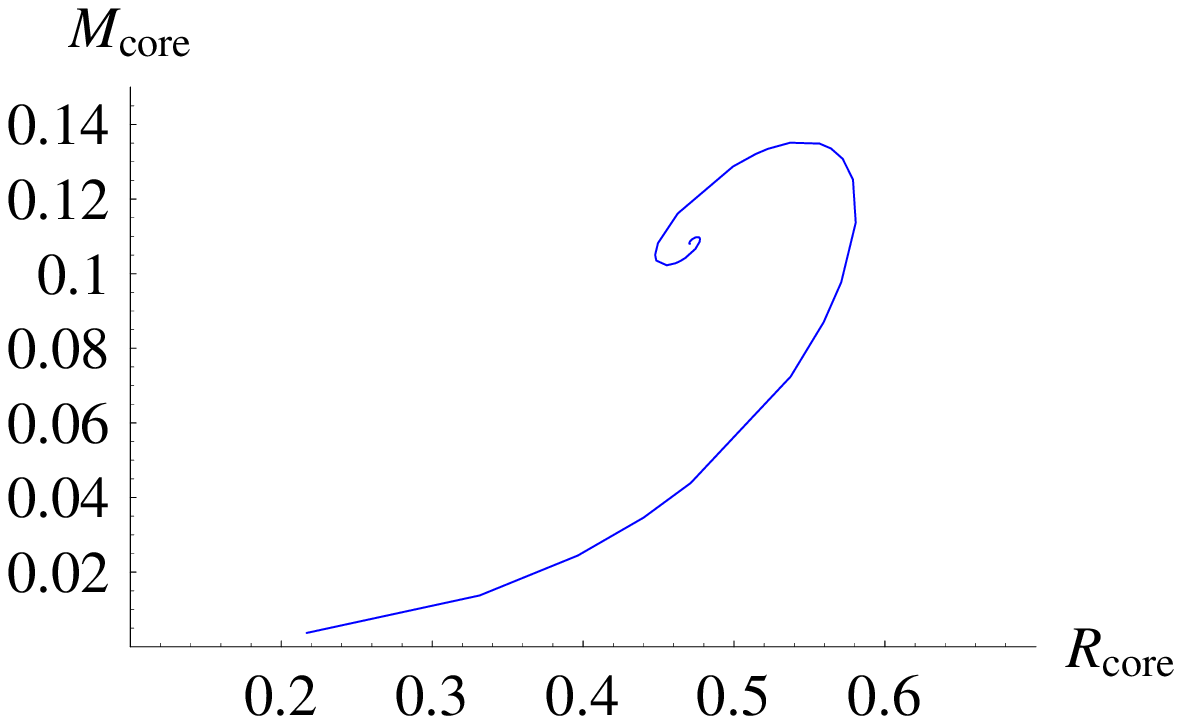} \hfill
\includegraphics[width=0.45\textwidth]{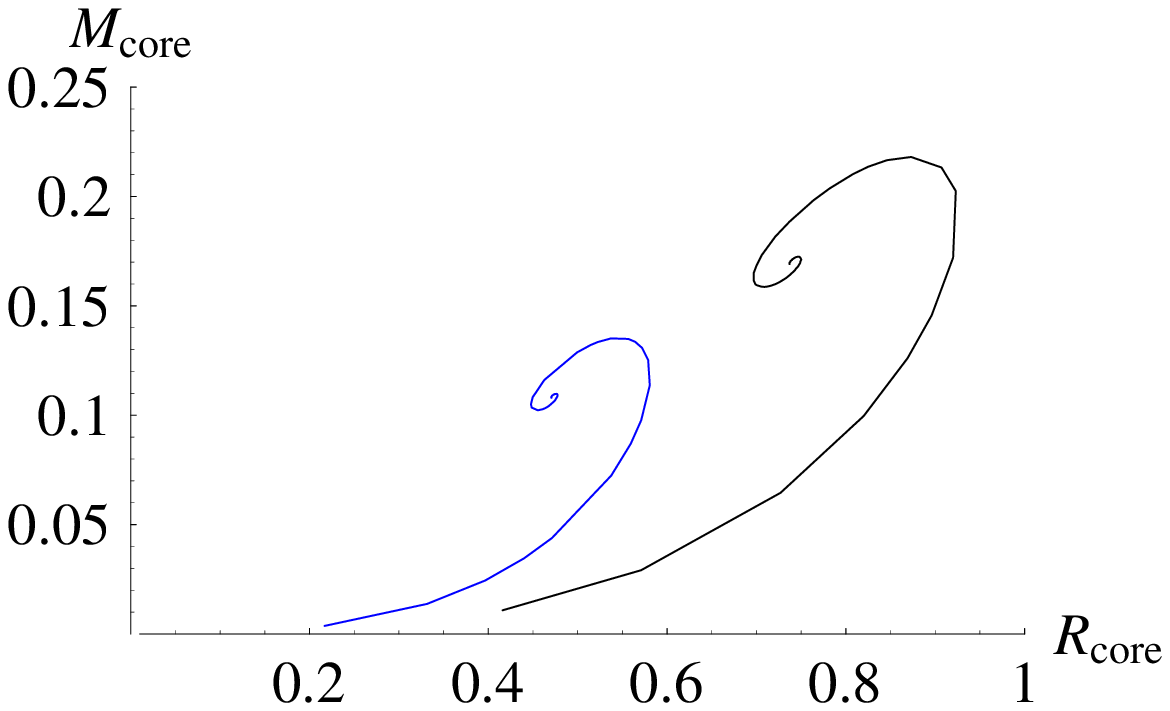} \\
\caption[mass and radius]{The relation between mass and radius of
the core of the multiquark star with (a) $n_{s}=0$, (b)
$n_{s}=0$~(blue) and $n_{s}=0.3$~(black). } \label{fig11}
\end{figure}

The baryon chemical potential distributions in the multiquark star
for $n_{s}=0,0.3$ are shown in Fig.~\ref{fig9.2}.  In the core
region, the chemical potential distributions of both cases are
similar due to the similarity of the equations of state for large
density.  A small jump of the chemical potential at the transition
radius between core and crust region is the artifact from the
power-law approximation.  The value of the chemical potential at
the transition radius from the full expression which we used in
the numerical simulations is slightly different from the
approximated value using the power-law.

The adiabatic index and sound speed of the multiquark phase for
$n_{s}=0.3$ are shown in Fig.~\ref{fig9.5}.  The adiabatic index
is higher than $n_{s}=0$ case but the sound speed in the low
density region is distinctively higher.  Around the transition
density, the sound speed reaches the maximum value of about 0.986
of the speed of light in vacuum.  For both $n_{s}=0,0.3$ cases, it
is obvious that the adiabatic index is closer to 1 in the core
reflecting the fact that the density distribution is more
condensed in the core region.  The adiabatic index reaches
$\lambda^{\prime}=2$ at the star surface since the the equation of
state at zero density is $P\propto \rho^{\lambda^{\prime}}$~(i.e.~
$\Gamma(\rho\to 0)=\lambda^{\prime}$ for Eqn.~(\ref{eq:rho})).

The {\it spiral} relation between mass and radius of the
multiquark star is shown in Fig.~\ref{fig10}.  As the central
density is increasing, the mass and radius of the $n_{s}=0~(0.3)$
multiquark star converge to the value of 0.659~(0.440) and
3.132~(1.704) respectively.  For the core, the mass and radius of
the core for $n_{s}=0~(0.3)$ converge to the value of
0.108~(0.169) and 0.471~(0.737).

Finally, we would like to estimate these limits of mass and radius
in the physical units.  Since our dimensionless quantities are
related to the physical quantities through conversion factors
given in Table~\ref{tab}~(Appendix A), both physical mass and
radius vary with the energy density of the nuclear matter phase as
$\propto 1/\sqrt{\text{energy density scale}}$.  For a multiquark
nuclear phase with energy density scale $10~\text{GeV/fm}^{3}$,
the conversion factor of the mass and radius are $5.91 M_{solar}$
and $8.71$ km respectively.  This would correspond to the
converging mass and radius~(in the limit of very large central
density) of $3.89~(2.60) M_{solar}$ and $27.29~(14.85)$ km for
$n_{s}=0~(0.3)$ multiquark star respectively.

In realistic situation, the nuclear phase in the outer region
could lose heat out to the space in the form of radiation.  The
nuclear matter in the outer region of the crust will cool down and
mostly become confined into neutrons and hadrons~(e.g. hyperons,
pions).  This would make the multiquark crust to end at shorter
radius than the estimated value and render the multiquark star to
be smaller and less massive than the estimated values in the
hypothetical prototype.  For example, for the energy density scale
$10~\text{GeV/fm}^{3}$, the critical density is $\rho_{c}\approx
1.5\times 10^{18}$ kg/m$^{3}~(n_{s}=0)$.  This is still a
sufficiently large density for the neutron layer to be formed.  If
the temperature of the nuclear matter in the crust region falls
below the deconfinement temperature, the multiquarks will be
confined into extremely dense neutrons and hadrons instead.  For a
typical neutron star, the distance of the neutron layer out to the
star surface is roughly 5-6 km~\cite{web}.  If we add this number
to the radius of the multiquark core, $0.471\times 8.71\simeq
4.10$ km, we end up with a more realistic estimation for the
multiquark star with radius $\sim 10$ km.  Regardless of the name,
only the core region is in the deconfined multiquark phase and the
content of the outer layers are the confined nucleons.

\section{Discussions and conclusion}

In the gluon-deconfined phase of the general Sakai-Sugimoto model,
multiquark states can exist in the intermediate temperatures below
the chiral symmetry restoration temperature provided that the
density is sufficiently large.  They are stable and preferred
thermodynamically over other phases and thus they can play an
important role in the physics of compact warm stars.  By analytic
and numerical methods, we demonstrate that the equation of state
of the multiquark nuclear matter can be approximated by two
power-laws in the small and large density region.  Roughly
speaking, the pressure is proportional to $d^{2}$ and $d^{7/5}$
for the small and large number density~($d$) regions respectively.

It is also found that the effect of the colour charges of the
multiquark is to reduce the pressure of the multiquarks when the
density is small.  At higher densities, multiquarks with colour
charges exert slightly larger pressure than the colourless ones.
 The temperature dependence of the entropy density shows an
$s\propto T^{5}$ relation and the colour charge dependence
$s_{colour}\propto n_{s}T$~(see Fig.~\ref{fig6} and
Eqn.~(\ref{eq:entropy})).  This implies that the multiquarks with
colour charges have larger entropy but their number of degrees of
freedom depend less sensitively on the temperature.  Multiquarks
in the deconfined phase behave like quasi-particles with the
entropy density being less sensitive to the temperature than the
gas of mostly free gluons and quarks in the $\chi_S$-QGP phase.

Using the power-law equations of state for both small and large
density regions, a spherically symmetric Einstein field equation
is solved to obtain the Tolman-Oppenheimer-Volkoff equation.  By
solving this equation numerically, we establish the mass, density
and pressure distribution of the hypothetical multiquark star.  It
turns out that the multiquark star is separated into two layers, a
core with higher density and a crust with lower density. Mass
limit curve is also obtained as well as the mass sequence plot
between the mass and radius of the multiquark star.  They show
typical spiral behaviour of the star sequence plots.  The mass
limit curve shows two peaks corresponding to the equation of state
of the small and the large density.  Analyses show that the most
contribution to the total mass is mainly from the crust.  The
adiabatic index at constant entropy, $\Gamma$, and the sound
speed, $c_{s}$, of the multiquark nuclear phase within the star
are calculated numerically.  For large density, $\Gamma$ is
approximately close to 1 and $c_{s}$ is roughly within range
$0.6-0.7$ of the speed of light.  For small density, $\Gamma$ is
in the range $1.3-2.0~(2.0-3.0)$ and $c_{s}$ is roughly
$0-0.85~(0-0.99)$ for multiquark with $n_{s}=0~(0.3)$.

\section*{Acknowledgments}
\indent P.B. is supported in part by the Thailand Research Fund
(TRF) and the Commission on Higher Education (CHE) under grant
MRG5180227. E.H. is supported by the Commission on Higher
Education~(CHE), Thailand under the program Strategic Scholarships
for Frontier Research Network for the Ph.D. Program Thai Doctoral
Degree for this research.

\appendix
\section{Dimensional translation table}
\begin{table}[ht]
\centering
\begin{tabular}{| c | c | c |}
\hline
quantity & dimensionless variable & physical variable \\[0.5ex]
\hline
pressure & $P$ & $\frac{c^{4}}{Gr_{0}^{2}}P$ \\[0.5ex]
density & $\rho$ & $\frac{c^{2}}{Gr_{0}^{2}}\rho$ \\[0.5ex]
mass & $M$ & $\frac{r_{0}c^{2}}{G}M$ \\[0.5ex]
radius & $r$ & $r_{0}r$ \\[0.8ex] \hline
\end{tabular}
\caption{Dimensional translation table of relevant physical
quantities, $r_{0}\equiv \left(\frac{G\mathcal N}{c^{4}\tau V_{3}}
\right)^{-1/2}=\left(\frac{G}{c^{4}}(\text{energy density scale})
\right)^{-1/2}$.} \label{tab}
\end{table}


\newpage

\begin{thebibliography}{11}

\bibitem{kar}
F. Karsch, ``Lattice QCD at finite temperature,'' AIP Conf. Proc.
{\bf 631} (2003) 112.
\bibitem{mp}
Marco Panero, ``Thermodynamics of the QCD plasma and the large-N
limit,'' {\it Phys. Rev. Lett.} {\bf 103} (2009) 232001,
[arXiv:0907.3719].
\bibitem{shuryak_colored states2}
E. V. Shuryak, I. Zahed, ``Towards a theory of binary bound states
in the quark gluon plasma,'' {\it Phys. Rev.} {\bf D70} (2004)
054507, [arXiv:hep-ph/0403127].
\bibitem{shuryak_colored states1}
E. V. Shuryak, I. Zahed, ``Rethinking the properties of the quark
gluon plasma at $T \sim T_{c}$'', {\it Phys. Rev.} {\bf C70}
(2004) 021901, [arXiv:hep-ph/0307267].
\bibitem{shuryak_colored states3}
J. Liao, E.V. Shuryak, ``Polymer chains and baryons in a strongly
coupled quark-gluon plasma,''{\it Nuc. Phys.} {\bf A775} (2006)
224, [arXiv:hep-th/0508035].
\bibitem{maldacena}
J. M. Maldacena, ``The Large N Limit of Superconformal Field
Theories and Supergravity,'' {\it Adv. Theor. Math. Phys.} {\bf 2}
(1998) 231-252 [{\it Int. J. Theor. Phys.} {\bf 38} (1998)
1113-1133], [arXiv:hep-th/9711200].
\bibitem{maldacena2}
Juan M. Maldacena, ``Wilson loops in large N field theories,''
{\it Phys. Rev. Lett.} {\bf 80} (1998) 4859-4862,
[arXiv:hep-th/9803002].
\bibitem{ry}
Soo-Jong Rey, Jung-Tay Yee, ``Macroscopic strings as heavy quarks
in large N gauge theory and anti-de Sitter supergravity," {\it
Eur. Phys. J.} {\bf C 22} (2001) 379-394, [arXiv:hep-th/9803001].
\bibitem{rty}
Soo-Jong Rey, Stefan Theisen, Jung-Tay Yee, ``Wilson-Polyakov loop
at finite temperature in large N gauge theory and anti-de Sitter
supergravity," {\it Nucl. Phys.} {\bf B 527} (1998) 171-186,
[arXiv:hep-th/9803135].
\bibitem{witb}
E. Witten, ``Baryons and Branes in Anti-de Sitter Space,'' {\it
JHEP} {\bf 07} (1998) 006, [arXiv:hep-th/9805112].
\bibitem{gross&ooguri}
D. J. Gross and H. Ooguri, ``Aspects of large N gauge theory
dynamics as seen by string theory,'' {\it Phys. Rev.} {\bf D58}
(1998) 106002, [arXiv:hep-th/9805129].
\bibitem{BISY}
A. Brandhuber, N. Itzhaki, J. Sonnenschein and S. Yankielowicz,
``Baryon from supergravity,'' {\it JHEP} {\bf 07} (1998) 046,
[arXiv:hep-th/9806158].
\bibitem{gho1}
K. Ghoroku, M. Ishihara, A. Nakamura and F. Toyoda, ``Multi-Quark
Baryons and Color Screening at Finite Temperature,'' {\it Phys.\
Rev.} {\bf D79} (2009) 066009,
 [arXiv:0806.0195 [hep-th]].
\bibitem{gho2}
K. Ghoroku and M. Ishihara, ``Baryons with D5 Brane Vertex and
$k$-Quarks States,'' {\it Phys. Rev.} {\bf D77} (2008) 086003,
[arXiv:0801.4216 [hep-th]].
\bibitem{Car}
M.V. Carlucci, F. Giannuzzi, G. Nardulli, M. Pellicoro and S.
Stramaglia,  ``AdS-QCD quark-antiquark potential, meson spectrum
and tetraquarks,''{\it Eur.\ Phys.\ J.}  {\bf C57} (2008) 569,
[arXiv:0711.2014 [hep-ph]].
\bibitem{Wen}
W-Y. Wen, ``Multi-quark potential from AdS/QCD,''{\it Int.\ J.\
Mod.\ Phys.}  {\bf A23} (2008) 4533, [arXiv:0708.2123 [hep-th]].
\bibitem{bch}
P. Burikham, A. Chatrabhuti and E. Hirunsirisawat, ''Exotic
multi-quark states in the deconfined phase from gravity dual
models,'' {\it JHEP} {\bf 05} (2009) 006, [arXiv:0811.0243
[hep-ph]].
\bibitem{ss lowE} T. Sakai and S. Sugimoto, ``Low Energy
Hadron Physics in Holographic QCD,'' {\it Prog. Theor. Phys.} {\bf
113} (2005) 843, [arXiv:hep-th/0412141].
\bibitem{ss more}
T. Sakai and S. Sugimoto, ``More on a Holographic Dual of QCD,''
{\it Prog. Theor. Phys.} {\bf 114} (2005) 1083,
[arXiv:hep-th/0507073].
\bibitem{Aharony}
O. Aharony, J. Sonnenschein and S. Yankielowicz, ``A Holographic
Model of Deconfinement and Chiral Symmetry Restoration,'' {\it
Annals Phys.} {\bf 322} (2007) 1420, [arXiv:hep-th/0604161].
\bibitem{tov}
J.R. Oppenheimer and G.M. Volkoff, ``On Massive Neutron Cores,''
{\it Phys. Rev.} {\bf 55} (1939) 374, R.C. Tolman, ``Static
solutions of Einstein's field equations for spheres of
fluid,''{\it Phys. Rev.} {\bf 55} (1939) 364.
\bibitem{bll}
Oren Bergman, Gilad Lifschytz, Matthew Lippert, ``Holographic
Nuclear Physics," {\it JHEP} {\bf 11} (2007) 056,
[arXiv:hep-th/0708.0326].
\bibitem{Callan}
C. G. Callan, A. Guijosa and K. G. Savvidy, ``Baryons and String
Creation from the Fivebrane Worldvolume Action,'' {\it Nucl.
Phys.} B 547 (1999) 127 [arXiv:hep-th/9810092].
\bibitem{Guijosa}
C. G. Callan, A. Guijosa, K. G. Savvidy and O. Tafjord, ``Baryons
and Flux Tubes in Confining Gauge Theories from Brane Actions,''
{\it Nucl. Phys.} B 555 (1999) 183
[arXiv:hep-th/9902197].
\bibitem{Tanii} N. Horigome and Y. Tanii,
``Holographic chiral phase transition with chemical potential,''
{\it JHEP} {\bf 01} (2007) 072, [arXiv:hep-th/0608198].
\bibitem{Kim} K. Y.
Kim, S. J. Sin and I. Zahed, ``Dense hadronic matter in
holographic QCD,'' [arXiv:hep-th/0608046].
\bibitem{apr}
A. Akmal, V.R. Pandharipande, and D.G. Ravenhall, ``The equation
of state of nucleon matter and neutron star structure", {\it
Phys.\ Rev.} {\bf C58} (1998) 1804, [arXiv:nucl-th/9804027].
\bibitem{ksz}
Keun-Young Kim, Sang-Jin Sin, Ismail Zahed, ``The Chiral Model of
Sakai-Sugimoto at Finite Baryon Density", {\it JHEP} {\bf 01}
(2008) 002, [arXiv:0708.1469 [hep-th]].
\bibitem{holo_n_star}
J. de Boer, K. Papadodimas and E. Verlinde, ``Holographic Neutron
Stars,'' [arXiv:0907.2695 [hep-th]].
\bibitem{web}
F. Weber, ``Strange Quark Matter and Compact Stars", {\it Prog.\
Part.\ Nucl.\ Phys.}  {\bf 54} (2005) 193,
[arXiv:astro-ph/0407155].

\end{thebibliography}
\end{document}